\documentclass[english]{iopart}

\usepackage[dvips]{graphicx}

\begin{document}

\title[Strange attractor and PC]
{\Large A birational mapping with a strange attractor:
Post critical set and covariant curves \\ \quad 14th June 2009}

\author{ 
M. Bouamra$^\S$, S. Hassani$^\S$ and
J.-M. Maillard$^\ddag$}
\address{\S  Centre de Recherche Nucl\'eaire d'Alger, \\
2 Bd. Frantz Fanon, BP 399, 16000 Alger, Algeria}
\address{\ddag\ LPTMC, CNRS, Universit\'e de Paris, Tour 24,
 4\`eme \'etage, case 121, \\
 4 Place Jussieu, 75252 Paris Cedex 05, France} 
\ead{maillard@lptmc.jussieu.fr, maillard@lptl.jussieu.fr, bouamrafr@yahoo.com}
 
\begin{abstract}
We consider some two-dimensional birational transformations.
One of them is a  birational deformation of the H\'enon map.
For some of these birational mappings, the post critical set (i.e.
the iterates of the critical set) is infinite and we show that this
gives straightforwardly the algebraic covariant curves of the 
transformation when they exist.
These covariant curves are used to build the preserved meromorphic
two-form.
One may have also an infinite post critical set yielding a
covariant curve which is not algebraic (transcendent).
For two of the birational mappings considered, the post critical set is 
not infinite and we claim
that there is no algebraic covariant curve and
 no preserved meromorphic two-form.
For these two mappings with non infinite post critical sets,
attracting sets occur and we show that they pass the usual tests
(Lyapunov exponents and the fractal dimension)
for being strange attractors. The strange attractor of one of these two
mappings is unbounded.
\end{abstract}


\noindent {\bf AMS Classification scheme numbers}: 34M55, 
47E05, 81Qxx, 32G34, 34Lxx, 34Mxx, 14Kxx


\section{Introduction}
\label{intro}

The study of dynamical systems uses the notion of sensitivity to initial
conditions as a criterion of the chaotic behavior. 
A large set  of the results in the theory of dynamical systems have
 been proven for hyperbolic systems 
(sometimes with the introduction of symbolic dynamics).
Otherwise, the study of a chaotic mapping is performed along various 
phenomenological and/or probabilistic approaches. 
In this dominant approach of dynamical systems
 the focus is on the system seen as a transformation 
of {\em real} variables, the analysis being dominated
besides computer experiments,  by  
functional analysis and differential geometry.
The study consists of orbits generated
on computers, phase portraits, bifurcations analysis 
and computation of Lyapunov exponents. Had the system an attracting
set which is not a manifold, the fractal dimension is introduced.
This phenomenological and/or {\em probabilistic viewpoint}
corresponds to the  mainstream approach of dynamical systems.
Most of the examples studied in the literature correspond to
iteration of polynomial or rational mappings.
Another drastically different approach 
can be introduced and 
corresponds to an {\em algebraic and topological approach} 
of dynamical system.
The mapping is seen as a dynamical system of
{\em complex variables} (complex projective space) 
and studied in the framework~\cite{dil-fav-01,bed-kim-04,bed-kim-tru-ab-ma-08}
of a cohomology of curves in complex projective spaces.
In this {\em topological} viewpoint, one counts integers
 (fixed points, degrees),
one deals with singularities and
with {\em blow up of points} and  {\em blow down of curves}~\cite{mcmullen-07}.
The matching of these two drastically different descriptions 
of discrete dynamical systems is far from being a simple question.

Consider a two-dimensional {\em reversible} mapping $K$:
\begin{eqnarray}
K: \,\, \quad  (u,v) \quad 
\,\,\longrightarrow \,\, \qquad 
\Bigl( K_u(u,v), \, K_v(u,v) \Bigr) 
\end{eqnarray}
The components $K_u(u,v)$, $K_v(u,v)$ may be polynomials,
or they may be rational. Both components may be polynomials but the
inverse transformation $K^{-1}$ has rational components.

In studying the dynamics of a mapping having rational components,
one encounters quickly that the mapping is ill-defined as a continuous
one because of the existence of a finite set of indeterminacy points.
The {\em indeterminacy} set ${\cal I}(K)$ of mapping $K$ is the finite set
of points for which a component of $K(u,v)$ has the form $0/0$.
Polynomial mappings have, of course, no indeterminacy set.

The {\em critical set} consists of those algebraic 
varieties that cancel the Jacobian
$J[K](u,v)$ of the  mapping $K$. Including
 also the algebraic varieties such that
$J[K](u,v)=\, \infty$ introduces the 
{\em exceptional locus}. We denote both
of them by ${\cal E}(K)$.
Mappings with constant Jacobian have, of course, no critical set.

For {\em reversible} two-dimensional mappings, one
 may want to distinguish between
bi-polynomial\footnote[5]{Their inverse are also polynomial
 transformations.} transformations such as the
 H\'enon mapping~\cite{henon-76},
polynomial mappings that have a rational inverse, such those studied 
in~\cite{BiGaMi_IJBC99} from the point of view of bifurcations due
to contact of phase curves (basin of boundaries, saddles) with
the indeterminacy set and exceptional
 locus\footnote[2]{In~\cite{BiGaMi_IJBC99,bi-mi-ga-00},
 the notions of
set of non-definition, prefocal curve and focal point are used.
See also~\cite{bi-cu-ph-97}.}, and {\em birational} mappings.

For {\em birational} mappings {\em generally}, the 
iterates of ${\cal E}(K)$ are 
{\em not curves but blow-down into points}:
\begin{eqnarray}
\label{PCgen}
K^n ( {\cal E}(K) ) \quad  \longrightarrow \quad \quad (u_n,\,  v_n),
 \qquad  n=\, \, 1, \,2, \, \cdots
\end{eqnarray}
These points $(u_n,\,  v_n)$ form the
 {\em post critical set}~\cite{bo-bo-ha-ma-05} (that we denote PC).

Knowing the full orbit (\ref{PCgen}) may not be easy.
For instance, in~\cite{bo-ha-ma-03} the orbits $K^n ({\cal E}(K))$ have
simple closed expressions.
The orbits $K^n ({\cal E}(K))$ may have algebraic expressions with
exponentially growing degrees {\em in the parameters}~\cite{bo-bo-ha-ma-05}.
For these examples and generically for a birational mapping, the 
post-critical set~\cite{bo-bo-ha-ma-05} (PC) is ``long''
meaning that, as the iteration proceeds, an infinite set 
of new points $(u_n, v_n)$ are obtained.

The PC orbit {\em may also be ``short''} by which it is meant
 that, after a finite number of iterations, the
 point $(u_n, v_n)$ settles in a
fixed finite point, or in $(\infty, \, \infty)$,
 and does not leave it.

In the framework~\cite{dil-fav-01,bed-kim-04}
of a cohomology of curves in complex projective spaces,
Diller and Favre have presented a method~\cite{dil-fav-01}, that gives
the conditions on the parameters
for which the mapping gets a complexity~\cite{ab-an-bo-ma-99} lower
 than that of the generic case.
This method  amounts to matching the iterates
of ${\cal E}(K)$ to the points of ${\cal I}(K)$.
The conditions $\, K^n ( {\cal E}(K) ) \in {\cal I}(K)\, $
(or $\, K^{(-n)} ( {\cal E}(K^{-1}) ) \in {\cal I}(K^{-1})$) 
give the value of the parameter for which 
the mapping gets a lower complexity
 than that of the generic case~\cite{dil-fav-01,bed-kim-04,ab-an-bo-ma-99}.
In other words the {\em complexity reduction},
which breaks the analytically stable~\cite{dil-fav-01}
 character of the mapping,
 will correspond to situations where
some points of the orbit of the exceptional locus 
($ K^n ( {\cal E}(K) )$) encounter
the indeterminacy set  ${\cal I}(K)$.

By ``complexity'', it is meant many quantities.
When one considers the degree $\, d(n)$ 
of the numerators (or denominators)
of the successive $\, n$-th iterate
 by mapping $K$ of a rational expression,
the growth of this degree is (generically) exponential 
with $n$: $\, d(n) \sim \lambda^n$. The constant $\lambda$ has been 
called the ''growth-complexity''~\cite{ab-an-bo-ma-99}
 and for $\, CP_2$,
is closely related to the {\em Arnold complexity}~\cite{arnold-90,arnold-93}.
Let us also recall that two universal (or ''topological'') measures of 
the complexities were found to 
identify for many examples of
 birational transformations~\cite{ab-an-bo-ha-ma-00,ab-an-bo-ha-ma-99pl}, 
namely the previous (degree) 
growth-complexity~\cite{ab-an-bo-ma-99},  Arnold
 complexity~\cite{arnold-90,arnold-93,ab-an-bo-ha-ma-99pl,ab-an-bo-ha-ma-99pa},
and the (exponential of the) topological
entropy~\cite{ab-an-bo-ha-ma-00,ab-an-bo-ha-ma-99pl,ab-an-bo-ha-ma-99pa,ab-an-bo-ma-2000}. 
The topological entropy is related to the
 growth rate, for increasing $\, n$, of the number of fixed points of 
$\, K^n$~\cite{dil-fav-01,ab-an-bo-ha-ma-00,ab-an-bo-ha-ma-99pl,favre-98}.
For birational mappings it is given by the (roots of the) denominator of a 
{\em rational} generating function
through the dynamical zeta function~\cite{art-maz-65} 
\begin{eqnarray}
\zeta(t)\,  =\, \,\, \,
 \exp{ \left( \sum_{n=1}^{\infty}{\# {\rm fix}(K^n)}\cdot
 \frac{t^n}{n} \right) },
\end{eqnarray}
where $\, \, \#{\rm fix}(K^n) \, \, $ denotes the
 number of fixed points at order $n$.

All the examples we have studied are {\em birational
 mappings}~\cite{ab-an-bo-ha-ma-00,ab-an-bo-ha-ma-99pl,ab-an-bo-ma-2000,BoMaRo93c,BoMa95},
and  we encountered the seemingly
discrepancy for a mapping to have non-zero
(degree-growth~\cite{ab-an-bo-ma-2000, ab-an-bo-ma-99} or
Arnold growth rate~\cite{ab-an-bo-ha-ma-00}) complexity,
or topological entropy~\cite{ab-an-bo-ha-ma-99pl}, while the orbits
(almost) always {\em look like} curves having non-positive Lyapunov exponents.
The regions where the chaos~\cite{gumowski-80,afrajmovich-94,gonchenko-97}
(Smale's horseshoe, homoclinic tangles, ...) is ``hidden'',
should be concentrated in {\em extremely narrow regions}.
Note that Bedford and Diller~\cite{BedDill} showed, for the
mapping of~\cite{ab-an-bo-ha-ma-99pl,ab-an-bo-ha-ma-99pa},
how to build the
invariant measure corresponding to non-zero {\em positive} Lyapunov exponents,
which corresponds to a {\em very slim Cantor set}.
Note that this invariant {\em real-measure} is drastically different from
the  {\em complex measure meromorphic two-form} of the
 mapping.

Furthermore, we reported in a previous paper~\cite{bo-bo-ha-ma-05} on two
birational mappings presenting very similar characteristics as far as
topological concepts are concerned.
They share the same identification between Arnold 
complexity growth rate and the (exponential of the) topological
 entropy~\cite{ab-an-bo-ha-ma-99pl}. 
The complexity reduction corresponds to the same algebraic numbers 
given by the same family of polynomials with integer coefficients.
Otherwise, leaving the algebraic-topologic description,
 these two mappings show different behaviors on other aspects.
One mapping~\cite{bo-ha-ma-03} preserves a meromorphic 
two-form~\cite{{bo-bo-ha-ma-05}} in the 
whole parameter space, while the other~\cite{bo-bo-ha-ma-05} does not 
 have a preserved meromorphic two-form for generic values of the
parameters. However, on some selected algebraic subvarieties of
the parameter space, the second mapping has a meromorphic
two-form. We showed in this case that, the
 fixed points of the birational mapping $\, K$
are such that $\, J[K^n]=\, 1$, where $J[K^n]$ is 
the Jacobian of $\, K^n$ evaluated at the fixed point of $\, K^n$.
For those cases, where a meromorphic two-form has not been found, the values
of $J[K^n]$ for the fixed points of $\, K^n$
are different from 1. We concluded that {\em this
mapping has no meromorphic two-form, since if it had one, this two-form would have
to accomodate all these ``non standard''
 fixed points whose number is infinite}.

In addition, we have considered~\cite{bo-bo-ha-ma-05} the vizualization of
the iterates of arbitrary initial points
 showing structures which, though similar,
are not converging towards the {\em post-critical set}
that is the iterates of the critical set.
No conclusion was drawn on the nature of these structures.
In this respect, one recalls the paper
 by Bedford and Diller~\cite{bed-dil-05}
which discusses a criterion related to close
 approaches of the {\em post-critical
set} to the indeterminacy locus.

In this paper we focus on {\em birational mappings}, seizing 
the opportunity to use, for this specific class of transformations, 
the concept of {\em post-critical sets}~\cite{bo-bo-ha-ma-05} (PC), that
we show to be straightforwardly related to algebraic covariant curves and
preserved meromorphic two-forms when they exist.

We first recall some previously analyzed mappings ($K_1, K_2$ and $K_4$)
and one mapping $K_3$ taken from the literature, 
and show how to obtain, from the {\em post critical set}~\cite{bo-bo-ha-ma-05}, 
the (algebraic) covariant curves and the preserved meromorphic two-form.
This analysis can be performed on either the forward mapping or the
backward mapping. In both directions the post-critical set is long.

A natural question arises then on whether the post-critical set of a
birational mapping can be ``short'' in one direction and ``long'' in the other
direction. What kind of structures do we expect?
A birational mapping of this kind would be a good example to study
the matching between the two viewpoints (topological and probabilistic)
of description of discrete dynamical systems.
Our aim is an attempt to link the short/long aspect of the post-critical
set to the {\em forward invariant} set occurring in a polynomial
mapping with {\em strange attractor}~\cite{eck-rue-85}.

Unfortunately and to the best of our knowledge,
 most~\cite{an-va-st-ko-98} of the {\em strange
attractors}\footnote[1]{The literature on strange
 attractors is too large to be recalled here.
Strange attractors are usually described in terms
of periodic points and unstable manifolds, the
 genesis of the visible attractor
being visualized as some kind of random walk 
on the union of all periodic points~\cite{Cvita}. 
The relation between the strange attractors and other 
selected points of the large literature on chaos, the homoclinic 
and heteroclinic points is, to our knowledge, not 
a very clear one.}
in two-dimensional {\em invertible} mappings found in the literature
 are {\em polynomial} transformations.
This stems from the fact that it is common to consider an attracting set as
bounded (compact set).
In typical situation (neither necessary nor sufficient
condition~\cite{ruelle-06}),
these structures arise when a mapping stretches and folds an open set,
and maps its closure inside it. 
The unbounded chaotic trajectories
that occur naturally in birational mappings are thought
to be divergent orbits.

We want, here, to build a birational  (one-parameter)
 deformation of  polynomial mappings. The first mapping
 $H_d$ we introduce
 is a birational deformation of the
celebrated H\'enon map~\cite{henon-76}.
The deformed birational mapping depends on a further parameter $c$ which
when fixed to zero gives back the original H\'enon map.
This continuous deformation will show how the
 H\'enon {\em strange attractor}
is modified.
In the topological point of view, the deformed H\'enon map has the same
degree-complexity for generic values of $c$,
 while the strange attractor
{\em deforms continuously} and the 
{\em fractal dimension of the attractor varies} continuously
as a function of the deformation parameter $\, c$.
For this mapping, the post critical set is ``short'' in the forward direction
(and ``long'' in the backward direction). It has no covariant curve and 
 no preserved meromorphic two-form.

We introduce a second birational mapping $K$, which will show that the
boundedness is not required for the occurrence of an attracting chaotic
set\footnote[3]{Some authors mention the possibility, or the occurrence,
of unbounded attracting sets~\cite{bi-mi-ga-00,bro-chu-96}.}.
First, we will compute its degree-growth 
complexity~\cite{ab-an-bo-ma-99,BoMa95}
 and topological entropy~\cite{ab-an-bo-ha-ma-99pa}
to show that the mapping is actually chaotic.
The phase portraits of the mapping show an invariant structure.
We will show that these structures {\em pass the usual tests} commonly used
to characterize the {\em strange attractor} (positive Lyapunov exponent and
fractal dimension). These calculations are carried out even if the mapping
has unbounded orbits. Thanks to the simplicity of the mapping, the fixed
points (computed up to $n=\, 15$) are real. These fixed
points are all lying on the structure.
The post critical set of $K$ is also ``short'' (in
the forward direction).

This last mapping $K$ falls in a family
of maps of two-steps recurrences of linear fractional transformation
studied by Bedford and Kim~\cite{bed-kim-05} in terms of periodicities
and degree growth rate~\cite{ab-an-bo-ma-99}.
Periodicities in this type of recurrences have
 been studied in e.g.~\cite{cso-lac-01,gro-lad-05}.

The paper is organized as follows.
Section \ref{PC} deals with the computation 
of the {\em post critical sets}~\cite{bo-bo-ha-ma-05}
(PC) for some birational mappings. These mappings being previously published,
the aim is to show quickly the deep relation between the post critical
set and the covariant curves of these known mappings.
In section \ref{henondef}, we introduce a birationaly deformed H\'enon
map. Here also, we want to benefit from the much studied bipolynomial
H\'enon map to establish the effect of the short post critical set.
Section \ref{secondmap} presents the second two-dimensional birational
mapping that has also a short post critical set and for which
the Arnold complexity growth rate 
and the (exponential of the) topological entropy identify.
In section \ref{mappi}, from the analytical expressions of the Jacobian
at the fixed points of the mapping up to $K^{11}$, and the proliferation
of what we call ``non-standard fixed points''
 ($J[K^n] \, \ne 1$), we conclude,  on the
non existence of a preserved meromorphic two-form.
The phase portraits of the mapping show an attracting set,
section (\ref{ergo}) deals with an ergodic 
analysis. The Lyapunov exponents are computed
and the dimension of the attracting set is given by both
Kaplan-York conjecture and box-counting method.

\section{The post critical set and covariant curves}
\label{PC}

\subsection{The birational mapping $K_1$}
Consider the mapping $K_1$ analyzed\footnote[5]{The original mapping $K_1$
 was written in the variables $(1/u,\, 1/v)$.}
in~\cite{ab-an-bo-ha-ma-99pl,ab-an-bo-ha-ma-99pa,bo-ha-ma-97}:
\begin{eqnarray}
  K_1 : \quad (u, v)\, \,\quad 
\longrightarrow \,\,\quad (u', v') \, = \, \,  \,
   \Bigl({\frac {(u+1) \,v }{1-\epsilon u}} , \, \,  
{\frac{u}{1+u-\epsilon u}} \Bigr).  
\end{eqnarray}
Its Jacobian reads:
\begin{eqnarray}
 J[K_1](u,v) \, = \, \, \, \, 
 -{\frac {u+1}{(1-\epsilon u)(1+u-\epsilon u)^2}}.
\end{eqnarray}
Using the same terminology as in~\cite{dil-fav-01},
the critical set is given by:
\begin{eqnarray}
{\cal E}(K_1) \, = \,  \, 
\, \,\left \{ (u=-1);\, \, (u=1/\epsilon) ; \, \,
 \Bigl(u= {1 \over \epsilon -1} \Bigr) \right \}.  
\end{eqnarray}

The post critical set $K_1^n \left( {\cal E}(K_1) \right)$ is given by
\begin{eqnarray}
&&(-1, \, v) \quad  \longrightarrow \quad \quad
\Bigl( {\frac{1+(-1)^n}{n-2-n \epsilon}}, \, \, \,
 {\frac{1-(-1)^n}{n-1-(n+1) \epsilon}} \Bigr),
\nonumber  \\
&&(1/\epsilon, \, v) 
\quad \longrightarrow \quad\quad
 \Bigl ( {\frac{-1}{(n-1) \epsilon}}, \, \,  \,
{\frac{1}{1-(n-1) \epsilon}} \Bigr),
 \nonumber
\end{eqnarray}
and the orbit $K_1^n (u=1/(1-\epsilon))$ depends on $v$.

From the iterates of $(u=-1)$, one sees that an infinite
 number of points of the post critical set
lie on $u=\, 0$ or $v=\, 0$. The elimination of $n$
 in the iterates of $(u=\, 1/\epsilon)$
gives the algebraic curve $v-u+uv =\, 0$. {\em Such algebraic 
curves are actually covariant} under the action of the
 birational transformation.

Denoting $(u', \, v')\, = \, \, K_1(u, \, v)$, one verifies 
that the $\, K_1$-covariant polynomial
 $m_1(u,v)\, =\, \, u\, v\, (v-u+uv)$ 
is actually such that
\begin{eqnarray}
\label{prepreserv}
 {{m_1(u',v')} \over {m_1(u,v)}} \,\,  = \, \,\,\,   J[K_1](u,v),
\end{eqnarray}
and one immediately deduces~\cite{Quispel} that the corresponding
 meromorphic two-form
\begin{eqnarray}
\label{preserv}
 {{du \cdot dv } \over {m_1(u, \, v)}} \, \,\, = \, \, \, \, \, \, 
 {{du' \cdot dv' } \over {m_1(u', \, v')}}, 
\end{eqnarray}
is preserved by the birational transformation $\, K_1$. 

One  remarks that as
 $n \, \rightarrow\, \infty$,
 the orbit of the critical
set goes to $(0,\,0)$ which is fixed point of order one for $K_1$.

\subsection{The birational mapping $K_2$}
Consider now the birational mapping  $K_2$ 
analyzed in~\cite{bo-ha-ma-03} (see Eq. (9) in~\cite{bo-ha-ma-03}), 
with $c=\, 2-a-b$:
\begin{eqnarray}
&&  K_2 : \,\, (u, v)\, \quad \longrightarrow \, 
\quad  (u', v') \,\,\, = \,  \\
&& \qquad \quad \, = \,   \, 
\Bigl({\frac {a\, u v\, +(b-1)\cdot v\, +cu}
{(a-1) \cdot  uv \, + bv\, +cu}} , \, \, \, {\frac
{a\, uv\, +bv\, +(c-1) \cdot  u}
{(a-1) \cdot  uv\, +b \, v\, +c\, u}} \Bigr)
 \nonumber 
\end{eqnarray}
The Jacobian is
\begin{eqnarray}
 J[K_2](u,v) \, = \,\,\,\,\,
{\frac{u \, v}{((a-1) \cdot  u \, v \, + c \, u\, + \, b\, v)^3}}, 
\end{eqnarray}
and the exceptional locus reads
\begin{eqnarray}
{\cal E}(K_2) \,  = \, \, 
\, \,\left \{ (u=0);\,\, (v=0) ; \,\,
 \Bigl(v=\, {{- c \, u} 
\over { (a-1)\, u \, + b}} \Bigr) \right \}.  
\end{eqnarray}

The successive images of the critical
 set are (see Eq. (13) in~\cite{bo-ha-ma-03}):
\begin{eqnarray}
\label{go}
&&(0, \, v) \quad \longrightarrow \quad 
\Bigl( {{b-1 } \over {b}}\,, \, \, 1
\Bigr) \quad \longrightarrow \quad 
 \cdots \quad \longrightarrow \quad 
\Bigl( {{n\, (b-1) } \over {n\, b\, -(n-1)}}, \, 1\Bigr),
 \nonumber  \\
&&(u, \, 0) \quad \longrightarrow \quad
 \Bigl ( 1\, , \, \, {{c-1 } \over {c}}
\Bigr) \quad \longrightarrow  \quad 
\cdots \quad \longrightarrow  \quad
 \Bigl( 1, \; \; {{n (c-1)}\over{n c -(n-1)}} \Bigr), \nonumber \\
&& (u \, , \, {{- c \, u} 
\over { (a-1)\, u \, + b}}) \, \, \quad \longrightarrow \, \, \quad
 \Bigl( \infty \, , \,  \infty
\Bigr)  \, \, \quad \longrightarrow  \, \, \quad \cdots  \\
&& \qquad \qquad  \, \,  \longrightarrow  \, \,\quad 
 \Bigl( {{(n-1) a -(n-2)}\over{(n-1) (a-1)}} , \;\;
{{(n-1) a -(n-2)}\over{(n-1) (a-1)}} \Bigr). \nonumber
\end{eqnarray}
 From these iterates one notes that an infinite number of 
points of the post critical set lie
 respectively on $v=\, 1$, $u=\,1$ and
$u=\, v$. 
{\em These three lines are actually covariant}. 
 Introducing the $K_2$-preserved polynomial
 $m_2(u,v) =\,(u-1)(v-1)(u-v)$,  
one deduces from the relation~\cite{bo-ha-ma-03,Quispel} 
between the Jacobian of $\, K_2$
and the ratio of $m_2$ evaluated at $(u,v)$ and at $(u', \, v')$,
 its image by $\, K_2$ 
\begin{eqnarray}
\label{prepreserv2}
 {{m_2(u',v')} \over {m_2(u,v)}} \,\, =\, \,\, \,\,J[K_2](u,v), 
\end{eqnarray}
 a meromorphic two-form preserved by $\, K_2$.

Here also, the limit $n \rightarrow \infty$ of the iterates (\ref{go}) 
 go to $(1,1)$
which is a {\em fixed point} of order one for $K_2$.

\subsection{The birational mapping $K_3$}

We consider now another example taken from the
 analysis of strongly regular
graphs~\cite{jaeger-92}.
The birational mapping reads:
\begin{eqnarray}
\label{K3birat}
&& K_3: \,\, (u,v) \,  
\,\,\longrightarrow \,\,\,   (u', v') \, =\, \,\,
  \Bigl(1\, + {\frac{28(v-u)N_{uv}}{u\,D_{uv}}},\,
1\,  + {\frac{28\,(v-1) N_{uv}}{D_{uv} }} \Bigr), \nonumber \\
&& N_{uv}\, =\,\,  3\,uv\, +3\,u\, +v,  \\
\label{theK4}
&& D_{uv}\, =\,\, 
 -(c^2+35)\cdot uv^2\, +2\, (c^2+7)\cdot uv\, 
-28v^2\, -(c^2-49)\cdot u.  
 \nonumber 
\end{eqnarray}

The post critical set is infinite and the orbit is
 given in closed form again 
allowing to obtain algebraic covariant curves.
To have a preserved meromorphic two-form, a further covariant curve
needs considering the post-critical 
set of the {\em backward} mapping.
The calculations are slightly more tedious, but still simple,
and are detailed in \ref{K3}.
 One obtains the following  $K_3$-covariant polynomial, with
 a relation between the Jacobian of  $K_3$ and the ratio of 
this $K_3$-covariant polynomial
\begin{eqnarray}
&& m_3(u,v)\, =\, \,\, \, 
{\frac{ (u-1)(v-u)\cdot }{v-1}}
\cdot
 \left((c^2-49)\, v^2\, -2\,(c^2+49)\,v+c^2-49 \right), \nonumber \\
&& {\frac{ m_3(u',\, v')}{m_3(u,\, v)}}\,\, 
  = \,\,  \,  {\frac{ m_3(K_3(u,\, v))}{m_3(u,\, v)}}\,\,
  = \,  \, \,\,\, J[K_3](u, \, v).
    \nonumber
\end{eqnarray}
which, again, enables to deduce  the corresponding
 meromorphic two-form. 

\subsection{The birational mapping $K_4$, for $b=a$}
The fourth mapping $K_4$ is taken from~\cite{bo-bo-ha-ma-05}
(see Eq.(16) in~\cite{bo-bo-ha-ma-05}) and 
reads (with $c=\,2-a-b$)
\begin{eqnarray}
&& K_4: \, \quad \, (u,v) \quad
\,\,\longrightarrow \,\,  \\
&&  \quad \quad \quad \quad 
 \Bigl( {\frac{b\,  (v\, +1)\,u \,+(b-1)\, v} 
 {(a-1) \cdot  uv\, +a \cdot  (u+v)}},\,\,\,
{\frac{c \, (u \, +1)\, v \,  \, +(c-1) \, u}
{(a-1) \cdot  uv \, +a \cdot  (u+v) }} \Bigr), \nonumber
\end{eqnarray}
with Jacobian:
\begin{eqnarray}
J[K_4](u, \, v)\, =\,\, \, \,\,
 {\frac { (a+b+c-1)\cdot uv  }
{ ((a-1) \cdot u v\, +a \cdot (u+v))^{3}}}.
\end{eqnarray}

The exceptional varieties of the mapping are 
\begin{eqnarray}
&&{\cal E}(K_4) \, = \,\, \,  
\left \{ V_1,\, V_2, \,V_3 \right \}
\,  \,=\,\, \nonumber  \\
&&\qquad \,=\,\,
\left \{ (u=0);\,\,\,  (v=0);\, \, \,
 (u=\, {\frac{-a\, v}{a\, +\, (a-1)\cdot v}}) \right \}\, . 
\end{eqnarray}

For the parameters satisfying $b=\, a$,
the iterates $K_4^n(V_1)$  are given by
(see Appendix E in~\cite{bo-bo-ha-ma-05})
\begin{eqnarray}
&&K_4^n(V_1)\, =\, \, \Bigl(u_n, \,\, v_n \Bigr)
\qquad \hbox { with: } \qquad 
\sigma_1 = \, {\frac{3a^2-4a+2}{2(2a-1)}},  \nonumber \\
&&u_n \, = \,\, \,  
{\frac{2\, (2a-1)\, T_{n}(\sigma_1)\,
 +(5a-4)\,a \cdot U_{n-1}(\sigma_1)-2(2a-1)}
{2\, (2a-1)\,T_{n}(\sigma_1)\,
 +(5a-4)\, a \cdot U_{n-1}(\sigma_1)+2(2a-1) }},     
\nonumber \\
&&v_{2n} \, = \, \, \, 
{\frac{-2\, (2a-1)(5a-4)\cdot T_{n}(\sigma_1)\, 
 -3\, (3a-2)(a-2)\,a \cdot U_{n-1}(\sigma_1)}
{4\, (2a-1)^2\cdot T_{n}(\sigma_1)}},     \nonumber \\
&&v_{2n-1}\,  =  \,\, \,  \nonumber \\
&&\quad  \,\, \, {\frac{2 \, (2a-1)(a^2+2a-2) \cdot T_{n}(\sigma_1) \, 
 -(3a-2)(a-2)^2\, a \cdot U_{n-1}(\sigma_1)}
{-2 \,(2a-1)^2 \, a\cdot T_{n}(\sigma_1)\, 
 +\, (a-2)(3a-2)(2a-1)\,a \cdot U_{n-1}(\sigma_1)}},     
\nonumber
\end{eqnarray}
where $T_n, U_n$ are Chebyshev polynomials of order $n$ of, 
respectively, first and second kind. 

We have similar results for  the iterates $K_4^n(V_3)$:
\begin{eqnarray}
&&K_4^n(V_3)\, =\,   \,\, \Bigl(u_n, \,\, v_n \Bigr), 
\nonumber \\
&&u_n \, =\, \, \,
 {\frac{2\, (2a-1)\cdot T_{n}(\sigma_1)\,
+(3a-4)\,a \cdot U_{n-1}(\sigma_1)+2}
{2\,(2a-1) \cdot T_{n}(\sigma_1) \, 
+(3a-4)\,a \cdot U_{n-1}(\sigma_1)-2 }}, 
   \nonumber \\
&&v_{2n} \, =\, \, \,
 {\frac{-4\, (2a- 1) \cdot T_{n}(\sigma_1)\, 
-6 \, (a-1)\, a \cdot U_{n-1}(\sigma_1)}
{2\,(2a-1)\cdot T_{n}(\sigma_1)\, 
+3\, a^2 \cdot U_{n-1}(\sigma_1)}},
  \nonumber \\
&&v_{2n-1} \, =\, \, \,
 {\frac{-2\,(2a- 1) \cdot  T_{n}(\sigma_1)\, 
-(5a-4)\, a \cdot U_{n-1}(\sigma_1)}
{2\,(2a-1)\, a \cdot U_{n-1}(\sigma_1)}}.     \nonumber 
\end{eqnarray}

The iterates $K_4^n(V_2)$  read 
(with $\sigma_2 \,  =  \, \, (3a-4)/2$):
\begin{eqnarray}
\label{prev}
K_4^n(V_2)\, = \, \,\, \,
\Bigl(1, \,\,\,  {\frac{2\,  (2a-1)\cdot U_{n-1}(\sigma_2)}
{2\,  T_n(\sigma_2)\,\,  -(5a-4)\cdot U_{n-1}(\sigma_2) }} \Bigr).
 \nonumber
\end{eqnarray}

From the iterates $K_4^n(V_2)$ one sees that an infinite number
of points in the post critical set 
lie on the line $u =\, 1$ which, is {\em thus} 
covariant by the birational transformation $K_4$.
By inspection, one obtains that the orbits $K_4^n(V_1)$
and $K_4^n(V_3)$ are lying on
$( 2\, (2a-1)(u+v^2)\, +(5a-4) (1+u)\, v )\,=\,\, 0$,
and are then $K_4$-covariant,
yelding to introduce the degree three $K_4$-preserved polynomial
\begin{eqnarray}
m_4(u,v)\, =\,\,  (u-1) \cdot (2\, (2a-1)(u+v^2)\, +(5a-4) (1+u)\, v ).
\end{eqnarray}
From the relation between the Jacobian of $K_4$
and the $m_4$-ratio 
\begin{eqnarray}
 {{m_4(u',v')} \over {m_4(u,v)}}
 \,   \, =\, \, \,\,    \, J[K_4](u,v), 
\end{eqnarray}
one immediately deduces a meromorphic
 two-form $\, du\, dv/m_4(u, \, v)$.

Here also, we have for the three
 $K_4^n\left( {\cal E}(K_4) \right)$
at the limit $n \rightarrow \infty$ 
\begin{eqnarray}
&& \left\{
\begin{array}{ll}
u_n & \longrightarrow  \quad  1, \\ & \\
v_{n} & \longrightarrow  \quad 
{\frac{5a-4 \, \pm \sqrt{3(a-2)(3a-2)}}{2(1-2a)}},\,\,
\qquad (+:a>0, \,\,\,   -:a<0)
\end{array}\right. \\
&& \mbox{ for}\, \,  \,\,\,  
 a \,  \in \,\, \,    ]-\infty,\,   2/3]\, \, 
 \cup\, \,  [2,\,   \infty[,
\,\,\qquad  a \,  \ne\, \,   0,\,  \,  1/2,   \nonumber
\end{eqnarray}
which are fixed points of order one for the birational 
mapping $K_4$.

\subsection{The birational mapping $K_4$, for generic $a$ and $b$}

From the previous examples, one sees that the {\em post critical set}
is an infinite orbit which is given in closed form enabling an elimination of the
iteration index $n$, thus yielding an explicit expression
 for some algebraic covariant curves.
The cofactors associated with these algebraic covariants are such
that a relation like (\ref{prepreserv})
occurs allowing  a preserved meromorphic two-form to exist.
One remarks also that the accumulation of the 
post critical set is a fixed
point of the mapping.

We consider now the mapping $K_4$, but for generic values of the
parameters $a$ and $b$. For this case, the post critical
set for $v=\, 0$ 
(for generic values of $a$ and $b$ and along all
${\cal E}(K_4)$  see~\cite{bo-bo-ha-ma-05}),
 begins as
\begin{eqnarray}
\label{singsing}
&& (u_1, v_1) \,\,   =\, \, \, \,\, \,    
\Bigl( {\frac{b}{a}},\,\, {\frac{1-a-b}{a}}\Bigr), \nonumber \\
&& (u_2, v_2) \, =\, \, \,\,  \,  
\Bigl({\frac{(b-1)}{(a-1)}}{\frac{(C+b)}{(C+a)}},\,\,\,  
{\frac{(1-a-b)}{(a-1)}}
{\frac{(C-a-b)}{(C+a)}}\Bigr), \nonumber \\
&& (u_3, v_3) \, =\, \, \, \,    \,\,
\Bigl( {\frac{f(a,b)}{f(b,a)}}, \,\,\,   
{\frac{(1-a-b) g(a,b)}{C\, f(b,a)}},  
 \Bigr) \nonumber  \\
&& (u_4, v_4) \, =\,\,\, \,\,   \cdots \nonumber
\end{eqnarray}
with:
\begin{eqnarray}
&& f(a, b)\, =  \,\,\, (b-1) \cdot
 (C^2 \, -(a+3b) \cdot C \, +(3b^2+a-b-2)\, b),
 \nonumber \\
&& g(a, b)\, =  \,\, \,
C^3 \,-2\,  (a+b-2) \cdot C^2 \,-3 \,(a+b-2) \,ab \cdot  C \,-ab, 
\nonumber \\
&& C \, = \,\,\, (a^2+ab+b^2)\,-(a+b).  \nonumber
\end{eqnarray}

Do note that, in contrast with the situation encountered
in the previous examples, the degree growth of
(the numerator or denominator of) these $(u_n, v_n)$
{\em in the parameters} $a$ and $b$ is, now, {\em actually exponential, and,
thus, one does not expect algebraic closed forms for the successive iterates}
($u_n$, $v_n$).
Had these iterates a closed form, and if the elimination of the index $n$
from $u_n$ and $v_n$ were possible, this would
 have given a {\em non algebraic} covariant curve.
For this case, this {\em transcendantal} curve should simply shrink to 
$u-1 =\,0\, $ for $\, b=\, a$.

\subsection{Sum up}

For the previous examples of birational mappings,
we have straightforwardly obtained, from the post-critical set,
 the algebraic covariant curves, enabling in a second step to build
 the meromorphic two-forms preserved by the  mappings.
This has been possible for all cases where the orbits of the critical
set are obtained in closed forms.
We have found that this happens whenever the degree growth {\em in the
parameters for the iterates of the critical set} is polynomial.
This concept of "degree growth in the parameters of the PC" has been
already introduced in~\cite{bo-bo-ha-ma-05}, giving a strong relation
between the post critical set 
and meromorphic two-forms (when they exist). 

One may define the post-critical set (PC)
 as "integrable" when the degree growth {\em in the
parameters} of the orbits ({\em of the critical set}) is polynomial
and "non integrable" otherwise.
For the mappings $K_1$, $K_2$ $K_3$ and $K_4$ (for $b=a$), the corresponding
PC is integrable and the mappings have algebraic covariant curves.
For the mappings $K_4$ (generic $a$, $b$) and $K_5$ (see \ref{K5}), the
iterates of the critical set having an exponential degree growth in the
parameters (i.e. the PC is "non integrable"), we claim that there is no
algebraic covariant curves.

Since the covariance of curves (if any) should be valid in both
directions, the PC should be ``long'' in both directions.
Furthermore, when the "long" PC is integrable in one direction, it
should be integrable in the other direction.
This is the case for these examples (and other alike mappings).

The question on whether a PC can be ``short''
in one direction and ``long'' in the backward direction
is worthy to be considered. Can the corresponding birational
mapping have covariant curves?
If so, the correspondence, we have shown in our examples,
between the occurrence of algebraic covariant curves and ``long''
(and integrable) PC is just fortunate.
We may even imagine a "pathological" case where both PC are ``short''.
A birational mapping of this kind is given in \ref{K6}.

We mentioned in the introduction the {\em strange attractors} (in
their down-to-earth definition) which
arise usually in some polynomials mappings
where the {\em indeterminacy set} is empty and the critical
set and exceptional set are also empty (the Jacobian is a constant).

A natural question is then:  can an {\em algebraic and topological
concept such as the post-critical set}~\cite{bo-bo-ha-ma-05}
(``long'' or ``short'', ``integrable'' or non integrable)
be related to the structures known as
{\em strange attractors}? 
For this, we will recall the well known 
(bi-polynomial\footnote[2]{Its inverse is also a 
polynomial transformation.})
 H\'enon map~\cite{henon-76},
and we will deform it birationally.
Is the post-critical set of this deformed H\'enon map long in both
directions or is it ``long'' in one direction and ``short'' in the other
direction?
We introduce another simple birational mapping that happens to be
a sub-family of transformations considered by Bedford and Kim
in~\cite{bed-kim-05} and ask the same questions.

\section{Birational deformation of H\'enon mapping}
\label{henondef}

We take advantage of the much studied
 H\'enon map~\cite{henon-76}, $H$, to construct
a {\em birational mapping} with a non empty indeterminacy set.
The birational deformation, we introduce,
 should depend on a further parameter
in order to get back the original mapping.
To be as close as possible to the dynamics of $H$,
the birational deformation should not
add {\em real fixed points} of order one to the two of the original
H\'enon map. Note however that this
 constraint is not mandatory (see below).

Recall the classical H\'enon mapping~\cite{henon-76}
\begin{eqnarray}
\label{themapH}
H: \,\, (u,v)
\quad \,\,\longrightarrow \,\, \quad 
\Bigl( 1-a u^2+bv, \,\, u \Bigr),  
\end{eqnarray}
which, under the reversible transformation
\begin{eqnarray}
U: \,\, (u,v)
\quad \,\,\longrightarrow \,\, \quad 
\Bigl( {\frac{u}{1+c v}}, \,\, {\frac{v}{1+ c u}} \Bigr),  
\end{eqnarray}
becomes a deformed birational H\'enon mapping $\, H_d$,
\begin{eqnarray}
\label{themaph}
&& H_d \, = \,\, \, H \cdot U: \,\,           \\
&&    (u,v) \,\,\quad  \longrightarrow \,\, \quad 
\Bigl(    
1\, -a u^2\, +bv\, +U_1, \, \, \,  u\,\, - c\, {\frac{uv}{1+cv}} \Bigr),
 \nonumber \\
&& U_1 \,=\,\,\,
-c \cdot {\frac{u v}{(1+cu)(1+cv)^2 }}\, \,  \times
 \nonumber \\
&& \qquad \Bigl(
 (b\,v\, -a\,{u}^{2})\, c^{2} \cdot v 
- (2\,a{u}^{2}\,+a\,uv\,-2\,b\,v)\cdot c\,\,
-2\,a\,u\, +b \Bigl), \nonumber
\end{eqnarray}
with inverse:
\begin{eqnarray}
\label{themaphM}
&& H_d^{-1} \, = \,\,\,\,\,  U^{-1} \cdot H^{-1}.  \nonumber 
\end{eqnarray}

Note that the transformation $U$ is a homography
\begin{eqnarray}
u_n=\,\,
{\frac{ (u-v)(1+cu)^n \,  u }
 {(1+cu)^n(1+cv)\,  u \, \,  -(1+cv)^n(1+cu)\, v }},
\,\,\,\,\, v_n=\,\,u_n \left( u \leftrightarrow v \right), 
 \nonumber 
\end{eqnarray}
and that the deformed birational mapping $H_d$ reduces
 to the classical  H\'enon map for $c=\,0$.

There are four fixed points of order one for the mapping $H_d$
\begin{eqnarray}
&& u \, = \,\, \, (1\, +c v) \cdot  v, 
\qquad \hbox{with:}\nonumber \\
&& (a+c) \,  c^2 \cdot  v^4\, +(2c+a) \, c \, \cdot v^3\,
 -(c^2-2c-a) \cdot v^2\nonumber \\
&& \qquad \qquad \,-(c+b-1)\cdot v\, \, -1 \,=\,\, 0. \nonumber
\end{eqnarray}
For the usual values of the parameters $a=\,1.4$ and $b=\,0.3$,
 two fixed points
are complex for a large interval of the parameter $c$.

The Jacobian of mapping $H_d$ reads:
\begin{eqnarray}
J(H_d)\, = \, \,\,
-{\frac {  b \cdot  (1 \, + \,c\, (u+v))}{ (1\, +c\,v)^{2}
 \, (1+c\,u)^{2}}}. 
\end{eqnarray}

The indeterminacy set and exceptional 
locus of this birational mapping
are:
\begin{eqnarray}
&& {\cal I}(H_d)\, =   \,\,\,
\left \{ (0, -1/c),\,  \, (-1/c, \, 0),  \, \,
 (-1/c, \, -1/c),\, \, (\infty, \, \infty) \right \},  \\
&& {\cal E}(H_d) \, = \,\, \, \left \{ V_1,\, V_2,\, V_3  \right \} 
 \nonumber \\
&&\qquad  =\, 
\left \{ (u=\,-1/c), \, \,(v=\,-1/c),\,  \, 
\left( v= \,-(1+cu)/c \right)\right \}.
\end{eqnarray}

The post critical set of $H_d$ is then
\begin{eqnarray}
&& H_d(V_1) \quad  \longrightarrow \quad  
 \left( \infty, \, -1/c/(1+cv) \right)
\quad  \longrightarrow \quad \left( \infty,\, \infty \right),  
 \nonumber \\
&& H_d(V_2) \quad   \longrightarrow 
 \quad  \left( \infty,\, \infty \right),  
 \nonumber \\
&& H_d(V_3) \quad   \longrightarrow \quad 
  \left( (c^2-a-bc)/c^2,\, -1/c \right) \quad \longrightarrow \quad 
\left( \infty, \infty \right).  \nonumber 
\end{eqnarray}
One remarks that the orbits are ``short'' 
(ending at the fixed point $(\infty, \, \infty)$)
 and there are only two points
to benefit from the Diller-Favre criterion~\cite{dil-fav-01}.
For generic values of the parameters $a$, $b$ and $c$, the birational
mapping $H_d$ has a degree-growth of $\lambda=\,3$.
The match  of the critical set to the indeterminacy
set gives the conditions on the mapping where $H_d$ acquires less
complexity than $\lambda=\,3$. One finds $\lambda=\,2$ for $c=\,0$
corresponding to the classical H\'enon map. For $a=\, (c-b)\, c$, one has
 $\lambda =\, 2.414213 \, \cdots$ given
by (the absolute value of the inverse of the smallest root of)
$1-2t-t^2 =\,0$, and for $a=\, (c-b+1)\,c $, one obtains
 $\lambda =\,2.618033 \cdots$
given by $1 -3t +t^2 =\,0$.

\begin{figure}
\begin{center}
\includegraphics[scale=0.3,angle=-90]{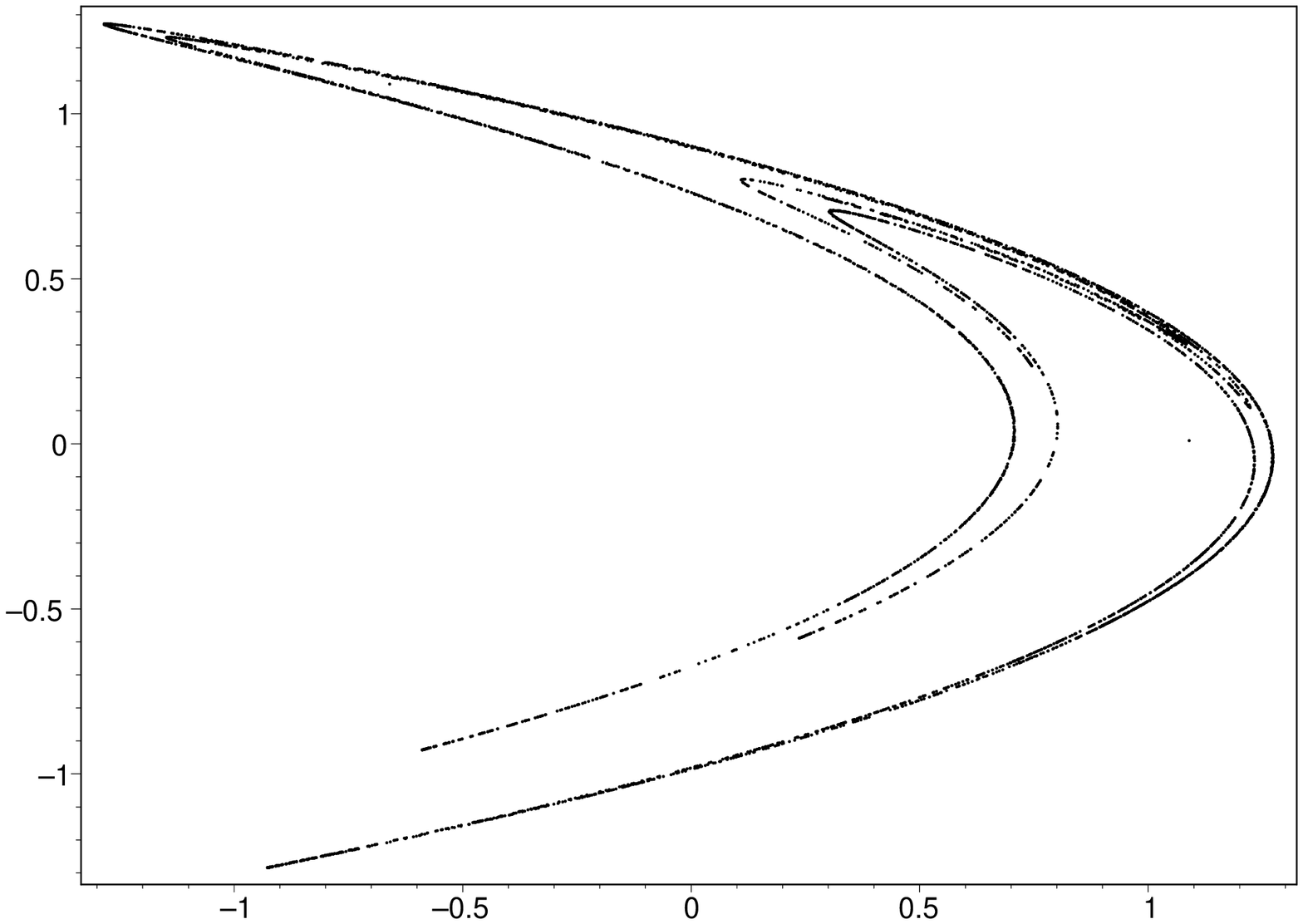}
\hfill
\includegraphics[scale=0.3,angle=-90]{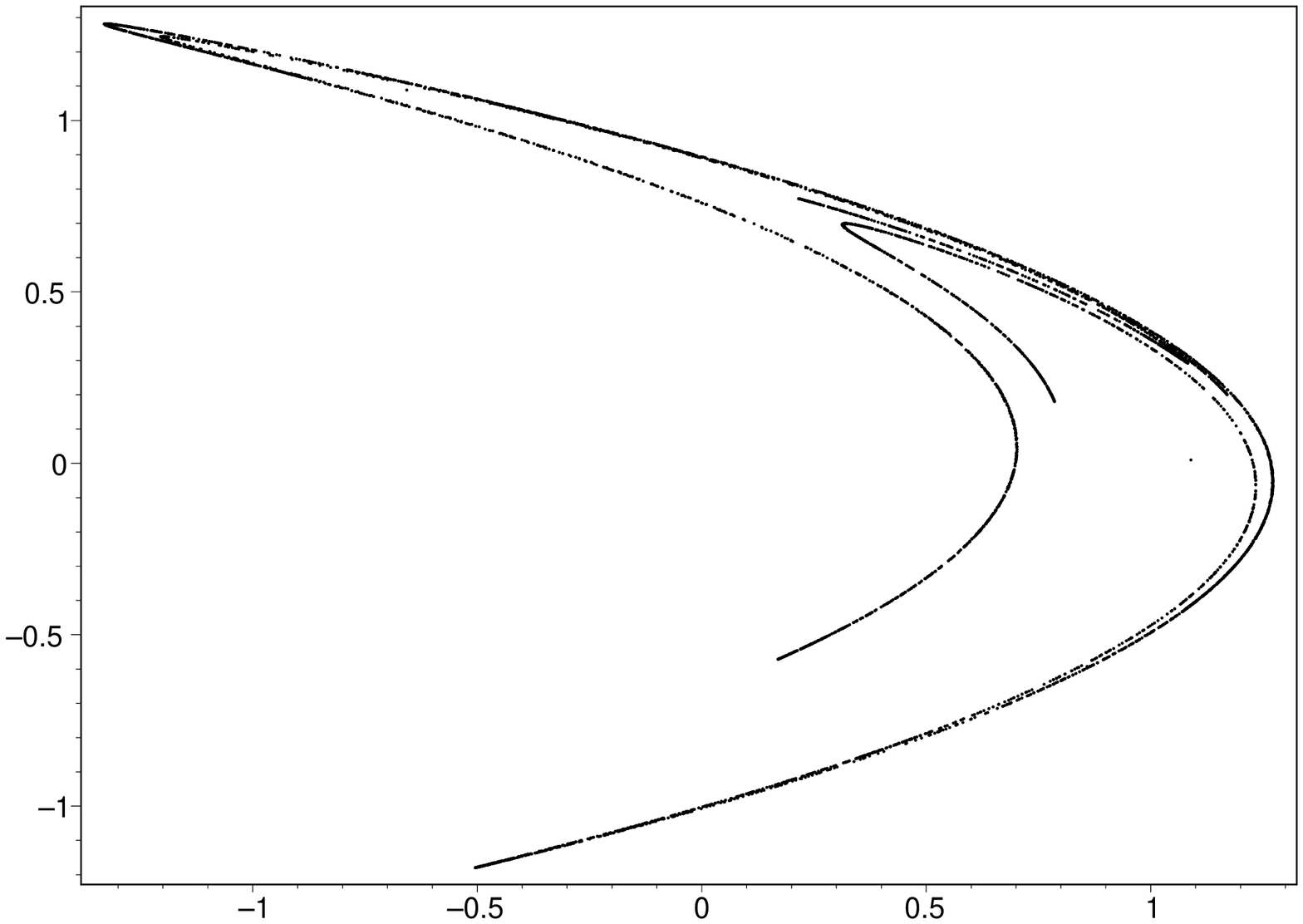}
\end{center}
\vskip .1cm
\protect\caption{The attracting sets for
 $c=\, 0$ (left) and $c=\,0.1$ (right)}
\label{figdefHenon1}
\end{figure}

\begin{figure}
\begin{center}
\includegraphics[scale=0.3,angle=-90]{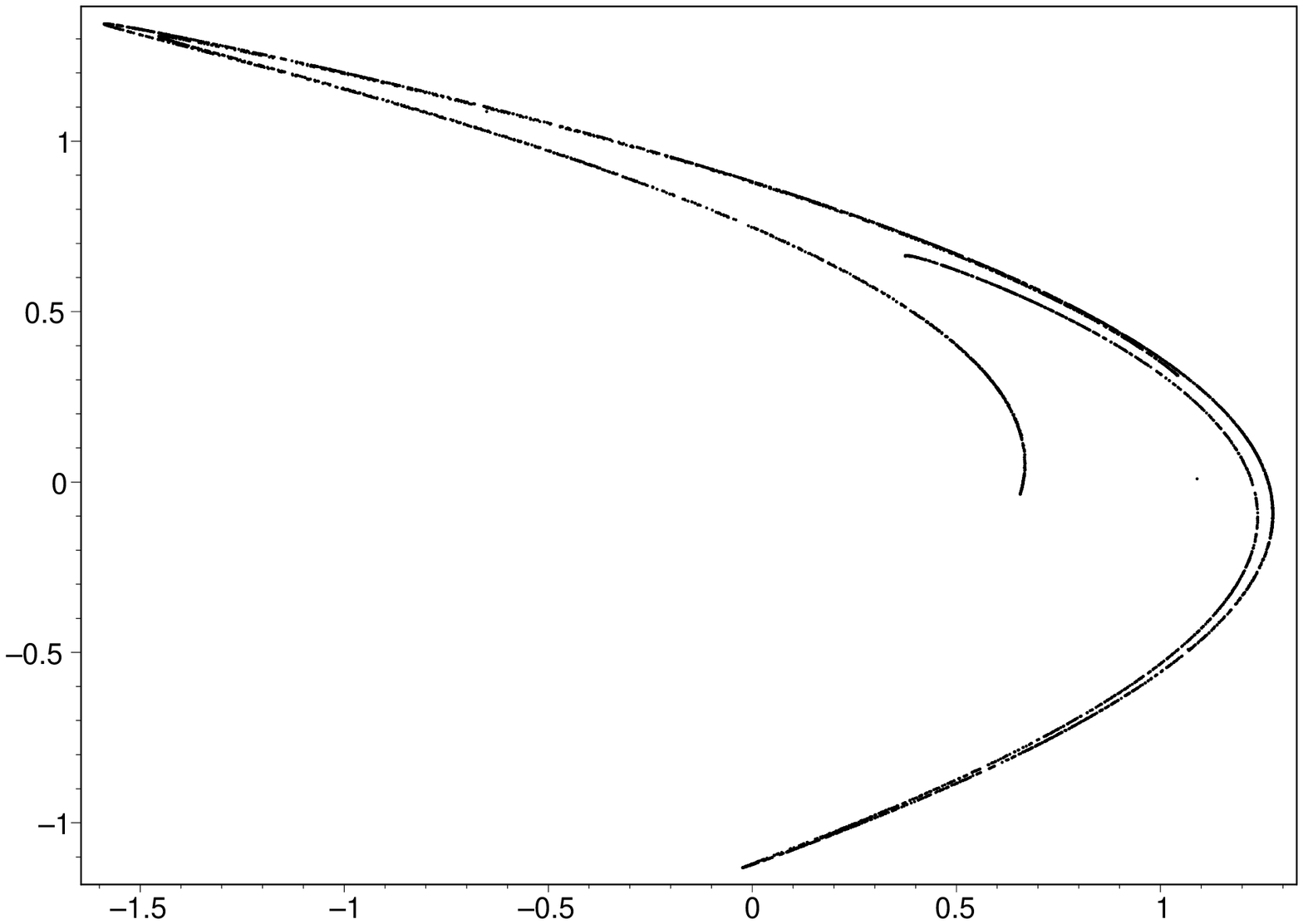}
\hfill
\includegraphics[scale=0.3,angle=-90]{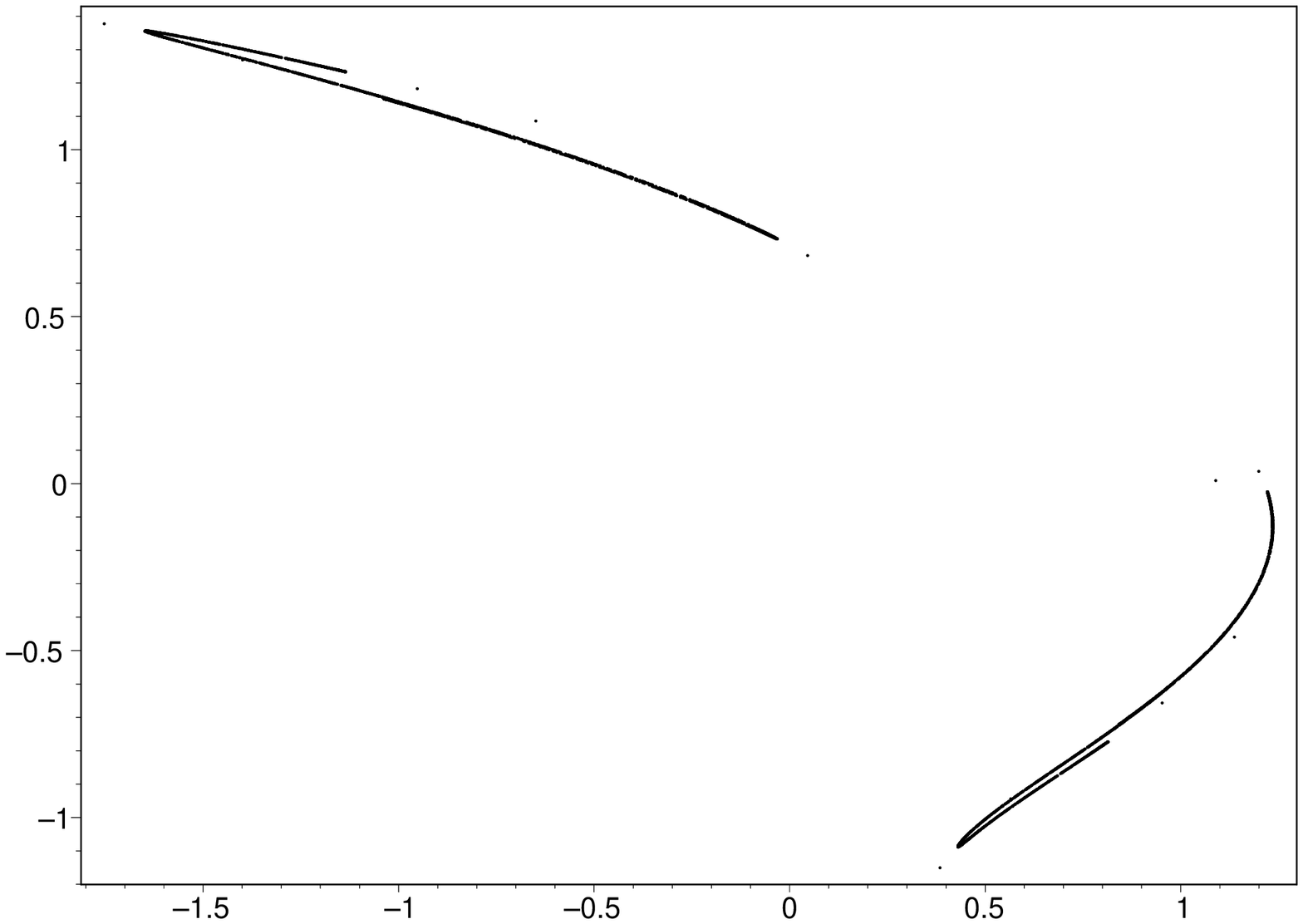}
\end{center}
\vskip .1cm
\protect\caption{The attracting sets
 for $c=0.3$ (left) and $c=0.38$ (right)}
\label{figdefHenon2}
\end{figure}

Remains to see what structures, the deformed H\'enon map gives.
Fixing the parameters $a=\, 1.4$, $b=\,0.3$ and for some values
of the parameter $c$, we give in Fig. \ref{figdefHenon1} and
Fig. \ref{figdefHenon2} the phase portraits of $H_d$.
The structures shown are reminiscent of the original H\'enon map
attractor.
The attractors shown have bassin of attraction outside of which
the iterates are unbounded.

\begin{figure}
\begin{center}
\includegraphics[scale=0.4,angle=-90]{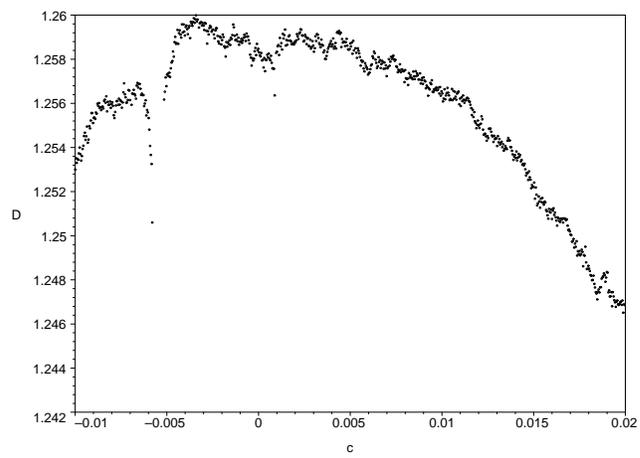}
\end{center}
\vskip .1cm
\protect\caption{Dimension of the attracting set
 for $H_d$, $(a=\,1.4, b=\,0.3)$ versus $c$.}
\label{figdefHenDim}
\end{figure}

Figure \ref{figdefHenDim} shows the {\em fractal 
dimension} (computed with the Lyapunov exponents, see (\ref{dimKY}))
 of the attracting set
for some values around $c=\, 0$.
We have clearly a {\em continuous deformation of the original H\'enon map}
as the parameter $c$ varies.

For the (backward) birational transformation 
 $H_d^{-1}$, there are three curves in the critical set
\begin{eqnarray}
&& {\cal E}(H_d^{-1}) \, = \,\,  \left \{ V_1, V_2, V_3  \right \} \\
&& =\,\left \{ (v=-1/c), (c u=\,c-b\,-acv^2), \,
\left( c^2\, v u\,=\,b+c^2v\,-ac^2\,v^3) \right)
\right \}.\nonumber 
\end{eqnarray}
Contrary to the forward mapping, the orbits of the critical set for
the backward  $H_d^{-1}$ are of infinite length
\begin{eqnarray}
\label{1overc}
&& H_d^{-1}(V_2)\quad  \longrightarrow \quad \left( 0,\, -1/c \right)
\quad  \longrightarrow \quad H_d^{-1}(V_1),   
\nonumber \\
&& H_d^{-1}(V_3)\quad  \longrightarrow \quad \left( \infty,\, \infty \right)\quad 
\longrightarrow \quad \left(-1/c,\, -1/c \right)
\quad  \longrightarrow \quad H_d^{-1}(V_1),  \nonumber \\
&& H_d^{-1}(V_1)\quad  \longrightarrow \quad \left(-1/c,\, 0 \right) 
\quad  \longrightarrow \quad  
\left( 0, \, -{1+c \over bc} \right) \, \, \, \, \, \cdots  \nonumber 
\end{eqnarray}
The (parameters) degree growth
 in the {\em  iterates of the critical set} is exponential $\lambda=\, 3$,
and thus we cannot have algebraic expressions associated
 with the post critical set (PC).

\begin{figure}
\begin{center}
\includegraphics[scale=0.4,angle=-90]{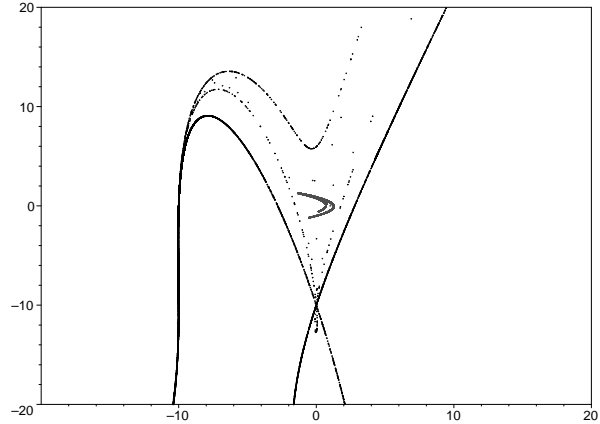}
\end{center}
\vskip .1cm
\protect\caption{Phase portrait of $H_d^{-1}$
 for $a=1.4, b=0.3, c=0.1$, together with
the strange attractor for $H_d$.}
\label{figdefHenBack}
\end{figure}

For the values\footnote[2]{To obtain the strange attractor for $H_d^{-1}$
equivalent to the attracting set for  $H_d$, the parameters should
be $(a,\, b) \rightarrow (a/b^2, \,1/b)$.} 
 $a=\,1.4$, $b=\,0.3$ and $c=\,0.1$, the 
phase portrait for the {\em backward}
 mapping is given in Fig. \ref{figdefHenBack}, where the {\em unbounded}
structure is obtained for any input point. 

{\bf Remark:}
Instead of the birational deformation of H\'enon map (\ref{themaph}),
we may consider the deformation
\begin{eqnarray}
\tilde{U}: \,\, \quad  (u,v) \quad 
\,\,\longrightarrow \,\, \quad 
\Bigl( {\frac{u}{1+c u}}, \,\, {\frac{v}{1+ c v}} \Bigr), 
\end{eqnarray}
for which the deformed birational mapping
 $H \cdot \tilde{U}$ will have a third real
fixed point. Here also, the birational
 mapping has a ``long'' post-critical set 
in the forward direction
and a ``short'' PC in the backward direction.
We still have a continuous deformation of the strange attractor
and phase portraits very similar to those shown
in Fig. \ref{figdefHenon1} and Fig. \ref{figdefHenon2} when the input
point is in the bassin of attraction.
For the input out of this bassin of attraction, the iterates
are attracted to the third fixed point, i.e. there are no divergent
(or unbounded) orbits.
We may note that this third fixed point of $H \cdot \tilde{U}$ goes
to infinity for $c=\,0$ (recall that for 
$c =\, 0$, one recovers the original
bi-polynomial H\'enon map).

In this example strange attractor points and points
of the post-critical set cannot overlap. That
 is very clear in the $c\, \rightarrow \, 0$ 
limit: points like $(0, \, -1/c)$
or  $(-1/c, \, 0)$ are quite
naturally  in the bassin of attraction
 of the super-attracting point
at infinity of the H\'enon mapping.

\section{A birational mapping: $K$}
\label{secondmap}

To build a birational mapping\footnote[1]{
This birational mapping is a slight modification
of a birational mapping considered
 by Bedford and Kim (Eq. (6.4) in \cite{bed-kim-05}). 
} $K$, we consider 
 a combination of the Cremona inverse
\begin{eqnarray}
\label{thej}
j: \,\,\quad  (x,\, y,\, z)\,\quad 
\,\,\longrightarrow \,\,\,\,  \quad 
\Bigl( yz,\, \,  xz,\,\,   xy \Bigr),  
\end{eqnarray}
and the linear transformation
\begin{eqnarray}
\label{theC}
C: \,\, \,\quad  (x, \,y, \,z) \quad 
\,\,\longrightarrow \,\, \,\, \quad 
\Bigl(y \,+b \,z, \, z, \, x \Bigr),  
\end{eqnarray}
giving
\begin{eqnarray}
K=\,  C \cdot j: \,\,\quad  (x,\,y,\,z) \quad 
\,\,\longrightarrow \,\, \,\, \quad 
\Bigl(  (z \,+b\,y) \cdot x, \, \, xy, \, \, yz \Bigr). 
\end{eqnarray}
In the inhomogeneous variables $u=\, x/z$,
 $v=\, y/z$, the birational mapping becomes
\begin{eqnarray}
\label{themap}
K: \,\,  \quad  (u,v) \quad 
\,\,\longrightarrow \,\,  \quad 
\Bigl( {\frac{u}{v}}  \,+ \, b\, u,\,  \, u \Bigr), 
\end{eqnarray}
with inverse
\begin{eqnarray}
K^{-1}: \,\, \quad  (u, v) \quad 
\,\,\longrightarrow \,\,  \quad \quad
\Bigl( v,\, -{\frac{v}{b\, v\, -u}} \Bigr), 
\end{eqnarray}
the linear transformation (\ref{theC})
 becoming the collineation
$(u, v) \, \rightarrow  \, ((b+v)/u , \,1/u)$.

The Jacobians are variables dependent and read
\begin{eqnarray}
J[K]\,  = \, \,\,  {\frac{u}{v^2}}, \quad \quad \quad 
 J[K^{-1}]\,  =\, \,\,   {\frac{v}{(b\, v\, -u)^2}}.
\end{eqnarray}

The indeterminacy set and the exceptional locus,
 for the birational mapping $K$, are
\begin{eqnarray}
&&{\cal I}(K)\, =   \,\,\,
 \left \{ (0, \, 0); \,\,\, 
\Bigl( \infty, -{{1}\over{b}} \Bigr) \right \}, \\
&&{\cal E}(K) \,  = \, \,\,  
\, \,\left \{ (u=0);\,\, (v=0)  \right \}  
\end{eqnarray}
and read  for mapping $K^{-1}$:
\begin{eqnarray}
&&{\cal I}(K^{-1})\, \, =\,    \,\,
 \left \{ (0, \, 0); \,\, ( \infty, \, \infty ) \right \}, \\
&&{\cal E}(K^{-1}) \,  = \, \, \,
\, \,\left \{ (u=bv);\,\, (v=0)  \right \}.  
\end{eqnarray}

The post critical set is the 
image by $K$ of the exceptional set:
\begin{eqnarray}
&&  K(u=0) \quad \longrightarrow \quad (0, \, 0)\quad 
 \longrightarrow \quad (\infty, \, 0)\quad 
 \longrightarrow \quad (\infty, \, \infty), 
 \nonumber \\
&&  K(v=0) \quad  \longrightarrow \quad (\infty, u)
\quad  \longrightarrow \quad (\infty, \, \infty). \nonumber 
\end{eqnarray}
Here also, the orbit of the critical set $K^n\left({\cal E}(K) \right)$
is ``short''. By Diller-Favre criterion~\cite{dil-fav-01},  only 
 $b =\, 0$ yields a complexity reduction, otherwise
the mapping is ``asymptotically stable'' (terminology introduced in~\cite{dil-fav-01}).
At the value $b=\, 0$, the whole plane becomes
a fixed point of order six which is easily seen from transformations
(\ref{thej}), (\ref{theC}) which are, respectively, of order two and three.
The fixed points of order one $(1,1)$, order two
$(-1/2 \mp i \,\sqrt{3}/2,\,\,  -1/2 \pm i\,\sqrt{3}/2)$ and of order three,
$(1,\, -1)$, $(-1,\,-1)$ and $(-1,\, 1)$ are still existing,
but any deviation from these exact values throws the trajectory in the
fixed point of order six.

To calculate the topological entropy~\cite{ab-an-bo-ha-ma-99pl} for the birational
mapping (\ref{themap}),
one counts the number of primitive
cycles of order $\, n$, for the generic case $b \ne 0$. They are
\begin{eqnarray}
 \#{\rm fix}(K^n) \,  \, =\, \, \, \, 
[1,\, 1,\, 1,\, 0,\, 1,\, 0,\, 1,\, 1,\, 1,\, 1, \,2,\, 2,\, 3,\, 3,\, 4,\, 5, \,\cdots\, ],
\end{eqnarray}
from which we infer the (rational) dynamical zeta function $\zeta(t)$:
\begin{eqnarray}
\label{generic}
\zeta(t)\,  =\, \, \,  \, {{ 1} \over {(1-t)\left( 1-t^2-t^3 \right)   }}. 
\end{eqnarray}
The absolute value of the inverse of the 
smallest root of $1-t^2-t^3=\, 0$ gives the
(exponential of the) topological entropy $h=\,  1.324717 \, \cdots$
This algebraic number is the {\em smallest
 Pisot\footnote[2]{On the occurrence of Pisot (and Salem numbers)
for degree growth complexity~\cite{ab-an-bo-ma-99,BoMa95} 
for birational transformations of two variables,
 see~\cite{valuative}.} number}~\cite{Pisot,Boyd,be-de-gr-pa-sc-92}. 

When one considers mapping (\ref{themap}) (in the 
homogeneous coordinates), the growth rate of 
the degrees of $x$ (or $y$ or $z$) is given
by the generating function
\begin{eqnarray}
 g(t)  \,  \, = \,  \, \,\, 
 {\frac { (2+2t+t^2) \cdot t}{ 1-t^2-t^3 }},  
\end{eqnarray}
and the degree growth complexity (absolute 
value of inverse of smallest root of the
denominator) gives again the smallest Pisot number
 $\lambda=\,1.324717\, \cdots \, $
This degree growth rate has been 
proven in~\cite{bed-kim-05} (and as a limiting
case in~\cite{cantat-99}).

We thus see, for this mapping (and similarly to the results
 obtained for other transformations previously
studied~\cite{bo-ha-ma-03,bo-bo-ha-ma-05,ab-an-bo-ha-ma-00,ab-an-bo-ha-ma-99pa}),
an identification  between the growth rate  of the number of fixed
points of $\, K^n$ and the growth rate of the
 degree~\cite{ab-an-bo-ma-99,BoMa95} of the iteration.

Following this criterion ($\lambda >1$, $h>1$), the birational mapping $K$
is chaotic.
However, this criterion
does not describe properly  the dynamics of the
mapping seen as a transformation in the {\em real plane}. 
It is based on the counting of degrees 
or fixed points irrespective of their stability.

The fixed point of order one is real for 
any real value of $b$. For $b=\, 1$, it
identifies with the fixed point at infinity.
For the mapping $K$, the fixed point of order one
is an unstable spiraling point for $b\,<\,0$, a 
stable spiraling for $0\,  <b\, <\, 3/4$,
a stable node for $3/4\,  <\, b<\, 1$, a 
saddle for $1<\, b\, <\, 3$ and an unstable node for $b \,>\, 3$. 
The fixed point of order 2 is real and saddle for $b<-1$ and
 for $b \,> \,3$.
 It {\em fuses} with the fixed point of
order one at $b= \,3$ and with
the "infinity" fixed point of order one at $b=\, -1$.
The fixed point of order three is real and saddle for any real $b$.

Note that, similarly to the mapping $H_d^{-1}$, the backward 
birational mapping
$K^{-1}$ has a ``long'' post-critical set. The iterates
 of the critical set has
also an {\em exponential degree growth in the parameter} $b$, ruling out
 the existence of algebraic covariant curves.

\section{The birational mapping $K$: phase portraits}
\label{mappi}

\begin{figure}
\begin{center}
\includegraphics[scale=0.5,angle=-90]{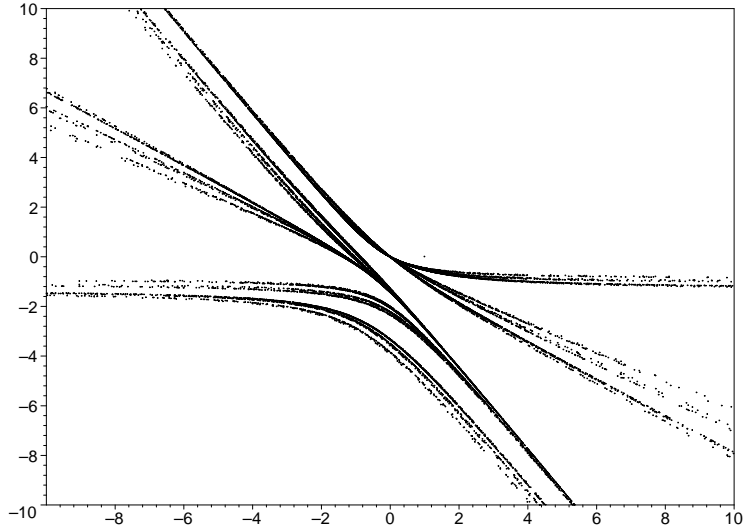}
\end{center}
\vskip .1cm
\protect\caption{Phase portrait in the variables $(u, v)$ for $b=\, -3/5$.}
\label{figPhaseuv}
\end{figure}

The phase portraits of the birational mapping $K$ show 
that for $ \vert b \, \vert  > \, 1$
the iterates are attracted to the fixed point 
at infinity. For $ 0 \, < b\,  <1$,
the fixed point of $K$ is an attractor.
For $ -1 < b <0$, the mapping is an attracting set.
We show in Fig. \ref{figPhaseuv}, the attractor 
obtained for the value $b=\, -3/5$.
This structure is independent of the initial
starting points of the iteration and looks very much like a set
of curves, a foliation of the $\, (u, \, v)$-plane.
The structure shown in Fig. \ref{figPhaseuv} is obtained from one
starting point.

\begin{figure}
\begin{center}
\includegraphics[scale=0.5,angle=-90]{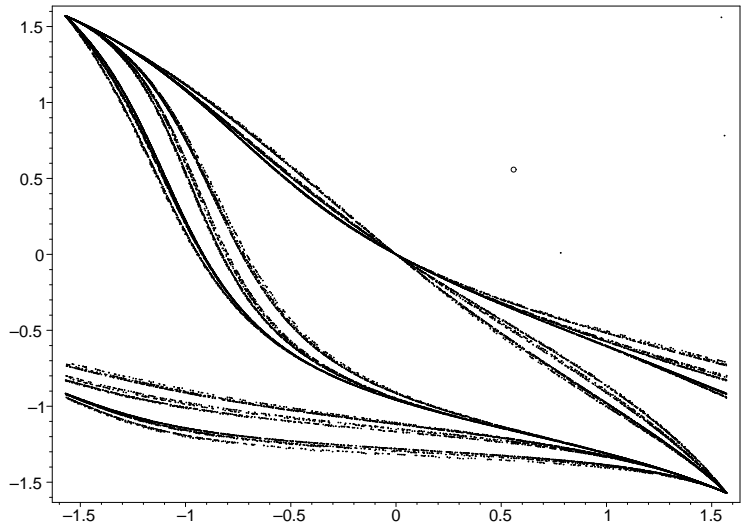}
\end{center}
\vskip .1cm
\protect\caption{Phase portrait in 
the variables $(\theta_u, \theta_v)$ for $b=\, -3/5$.
The open circle shows the fixed point of order one.}
\label{figPhaseuvC}
\end{figure}

This accumulation of curves has unbounded branches. A way to 
``see the global picture'' amounts to performing the plot 
in the variables~\cite{bo-bo-ha-ma-05} 
$\, \theta_u  \,= \, \arctan(u)$ and 
$\, \theta_v  \,= \, \arctan(v)$.
This way, the real plane is compactified to the box
$\,[ -\pi/2, \,\pi/2] \times [ -\pi/2, \, \pi/2]$.
Figure~\ref{figPhaseuvC} shows the phase portrait 
in the variables $(\theta_u, \theta_v)$.
The unbounded branches of Fig. \ref{figPhaseuv} 
appear in Fig.\ref{figPhaseuvC}
at the bunches $(\theta_u= \pm \pi/2,\,  \theta_v=\, \mp \pi/2)$ and
$(\theta_u= \pm \pi/2,\,  \theta_v= \, \theta_{\tilde{v}})$, 
the larger interval for $ \, \tilde{v}$ being $[b-1, \, b]$.
The attractor is thus not confined. This is consistent with the fact that
it is obtained for any initial point, the basin of attraction
 being the whole plane.

Since the birational mapping $K$ is not integrable
 according to the criterion $\lambda >1$
or $h > 1$, one may ask whether the structures,
shown in Fig.\ref{figPhaseuv} and Fig.\ref{figPhaseuvC}, are compatible
with a preserved meromorphic two-form or simply with covariant curves.
In fact the post critical set 
is ``short'', and there is no algebraic covariant
curve. In the following, we show another way to be convinced
on this non occurrence. 

\subsection{Non-standard fixed points}
\label{nonstand}
Denoting by $\, (u^{(n)}, \, v^{(n)})$, the image of a point $ (u,\, v)$,
by transformation $\, K^n$, 
the preservation of a two-form $m(u,v)$ means:
\begin{eqnarray}
&& {{du \cdot dv } \over {m(u, \, v)}} \, \,\, = \, \, \, \, \, \, 
  {{du^{(n)} \cdot dv^{(n)} } \over {m(u^{(n)}, \, v^{(n)})}}. 
\end{eqnarray}
If $ J[K^n](u, \, v)$ denotes the Jacobian of $\, K^n$, one has:
\begin{eqnarray}
J[K^n](u, \, v) \, \, = \, \, \, \,\,
 {{ m(u^{(n)}, \, v^{(n)})} \over { m(u, \, v)}} 
\, \, = \, \, \,\,\, {{ m(K^n(u, \, v))} \over { m(u, \, v)}}.
\end{eqnarray}
When evaluated at the fixed point $(u_f,\,  v_f)$ of $\, K^n$,
the Jacobian of $\, K^n$ is thus equal to +1.
This is what was obtained for many birational transformations we have
studied~\cite{bo-ha-ma-03,ab-an-bo-ha-ma-99pl,ab-an-bo-ha-ma-99pa}.
For some of these mappings, we evaluated
precisely a large number of $n$-cycles in order to 
get the dynamical zeta
 function~\cite{ab-an-bo-ha-ma-99pl,ab-an-bo-ha-ma-99pa}.
For all these mappings, a preserved meromorphic two-form exists.
However, we claimed for the mapping given in~\cite{bo-bo-ha-ma-05},
the non existence of a preserved meromorphic two-form
since the Jacobians evaluated at (a growing number of) the fixed points
of $K^n$ are no longer equal to one.
This mapping preserves a meromorphic two-form in some subspaces of the
parameters, and we showed, in this case, that the equality
$J[K^n](u_f, \, v_f)=\,1$ always holds, exception of a {\em finite number}
of fixed points.
Thus, even when a meromorphic two-form is preserved, the
equality $J[K^n](u_f, \, v_f)=\, 1$ 
{\em may be ruled out} for the fixed points of $\, K^n$
that {\em correspond to divisors of the two-form}.
Such ``non-standard'' fixed points of $\, K^n$ are such that
$\, m(u_f, \, v_f) \, = \, \, 0$ 
(or $\, m(u_f, \, v_f) \, = \, \, \infty$).

The number of these ``non-standard'' fixed
 points~\cite{bo-bo-ha-ma-05} of $\, K^n$ is
an indication of the degree of $m(u,v)$, if the latter exists.
When the number of such non-standard fixed points
becomes very large (infinite),
the existence of this algebraic two-form may be ruled out.

Thanks to the simplicity of the mappings of this paper,
the expressions of these Jacobians
evaluated at the fixed points may be obtained up to a large order.
For instance, at respectively the order 
$n=\,1$, $n=\,3$, $n=\,10$ and $n=\,11$,
they read ($(u_f, v_f)$ are the primitive fixed points of $K^n$)
\begin{eqnarray}
&&J[K]\left(u_f, v_f \right) \, =\,  \, 1-b, \quad \quad \,\,\,
J[K^3]\left(u_f, v_f \right) \, =\,  \, 1+b+b^2,  \nonumber \\
&&J[K^{10}]\left(u_f, v_f \right) \, =\,  \,
{\frac{ (1-b^{10})\, b^{10}}{(1+b)\cdot (1-b^5) \cdot P_1^{(10)} }},       \\
&&J[K^{11}]\left(u_f, v_f \right): \, \quad \, \,  \,\, \,
P_2^{(11)}\cdot J^2\, + P_1^{(11)}\cdot J +P_0^{(11)}\, =\,\, 0,     \nonumber  
\end{eqnarray}
with:
\begin{eqnarray}
&&P_1^{(10)}\, =\, \, b^8\,-4b^7\,+9b^6\,-15b^5\,
+16b^4\,-14b^3\,+8b^2\,-3b\,+1, \nonumber \\
&&P_2^{(11)}\, =\,\,  
\Bigl (-2+8\,b-22\,{b}^{2}+46\,{b}^{3}-72\,{b}^{4}+89\,{b}^{5}
-87\,b^{6}+68\,{b}^{7}  \nonumber \\
&&\qquad \qquad  -41\,{b}^{8}+19\,{b}^{9}
-6\,{b}^{10}+{b}^{11}\Bigr) \cdot  b, \nonumber \\
&&P_1^{(11)} \, =\, \, \,  
1-b+5\,{b}^{2}-2\,{b}^{3}+8\,{b}^{4}-8\,{b}^{5}
+14\,{b}^{6}-25\,{b}^{7}+16\,{b}^{8}  \nonumber \\
&&\qquad -35\,{b}^{9}+13\,{b}^{10}-38\,{b}^{11}+21\,{b}^{12} 
 -34\,{b}^{13}+22\,{b}^{14}-19\,{b}^{15}  \nonumber \\
&&\qquad +10\,{b}^{16}-5\,{b}^{17}+2\,{b}^{18}-{b}^{19}, \nonumber \\
&&P_0^{(11)} \, =\, \, \,   (1-b^{11})\, \, b^{15}/(1-b). \nonumber 
\end{eqnarray}

As far as the fixed points up to order eleven
 are sufficient to make a conclusion,
there is no value of the parameter $b$ that gives $J[K^n]$ equal
to unity for these  fixed points.
Considering only these fixed points, a preserved meromorphic
two-form for the  birational mapping $K$ should be, at least, of degree $134$.
In fact the proliferation of these non-standard fixed points
gives a firm indication on the {\em non existence 
of a preserved meromorphic
two-form}.

\section{The  birational mapping $K$: ergodic (probabilistic) analysis}
\label{ergo}

While the "complexity spectrum" of 
the mapping in~\cite{bo-bo-ha-ma-05} is
involved (see Figure 1 in~\cite{bo-bo-ha-ma-05}), the mapping of this
paper presents the same degree-growth complexity
 or (exponential of the) topological
entropy ($\lambda=\, h= \, 1.324717 \cdots$) 
for any value of the parameter $b \ne 0$.
Numerical investigation shows that the fixed points, up to the order
$n= \,15$, are real for real values of the parameter $b$.
We expect, then, to provide a clearer picture
on the ergodic aspect of the analysis.
We have seen~\cite{bo-bo-ha-ma-05} that the existence, or non existence, of
preserved meromorphic two-forms
 has (paradoxically)\footnote[2]{We have made in~\cite{bo-bo-ha-ma-05}
 a comparison between two sets of
birational transformations  exhibiting totally 
similar results as far as  {\em topological complexity} is concerned
(degree growth complexity, Arnold complexity  and  topological entropy),
but {\em drastically different numerical} results as far as 
Lyapunov exponents computation is concerned.} no impact 
on the topological complexity
but drastic consequences on the numerical computation of the
Lyapunov exponents.

If we believe the analysis of~\cite{bo-bo-ha-ma-05}, the {\em Lyapunov
exponents} should be zero in the case of a preserved meromorphic
two-form. This is then another check on whether the structure
shown in Fig. \ref{figPhaseuvC} correspond 
to a preserved meromorphic two-form.
We have computed the Lyapunov exponents for the mapping $K$ 
for the large values of $n$ as
\begin{eqnarray}
\sigma_1 \,=\,\,
 {1 \over n}\, \ln \left( \vert \lambda_1 \vert \right), \qquad
\sigma_2 \,=\, \,
{1 \over n}\, \ln \left( \vert \lambda_2 \vert \right), \qquad
\end{eqnarray}
where $\lambda_{1,2}$ are the eigenvalues of
\begin{eqnarray}
M^{(n)}\, =\,\,\, \, M(u^{(n-1)},\,  v^{(n-1)})
\,\, \, \, \cdots \,\,\,  \,
M(u^{(1)},\,\,   v^{(1)})\,   \cdot \,   M(u, \, v), 
\end{eqnarray}
$M$ being the tangent map evaluated at each point
$\left(u^{(n)}, v^{(n)}  \right)$
$=\, K^n \left(u, v \right)$.

\begin{figure}
\begin{center}
\includegraphics[scale=0.6,angle=0]{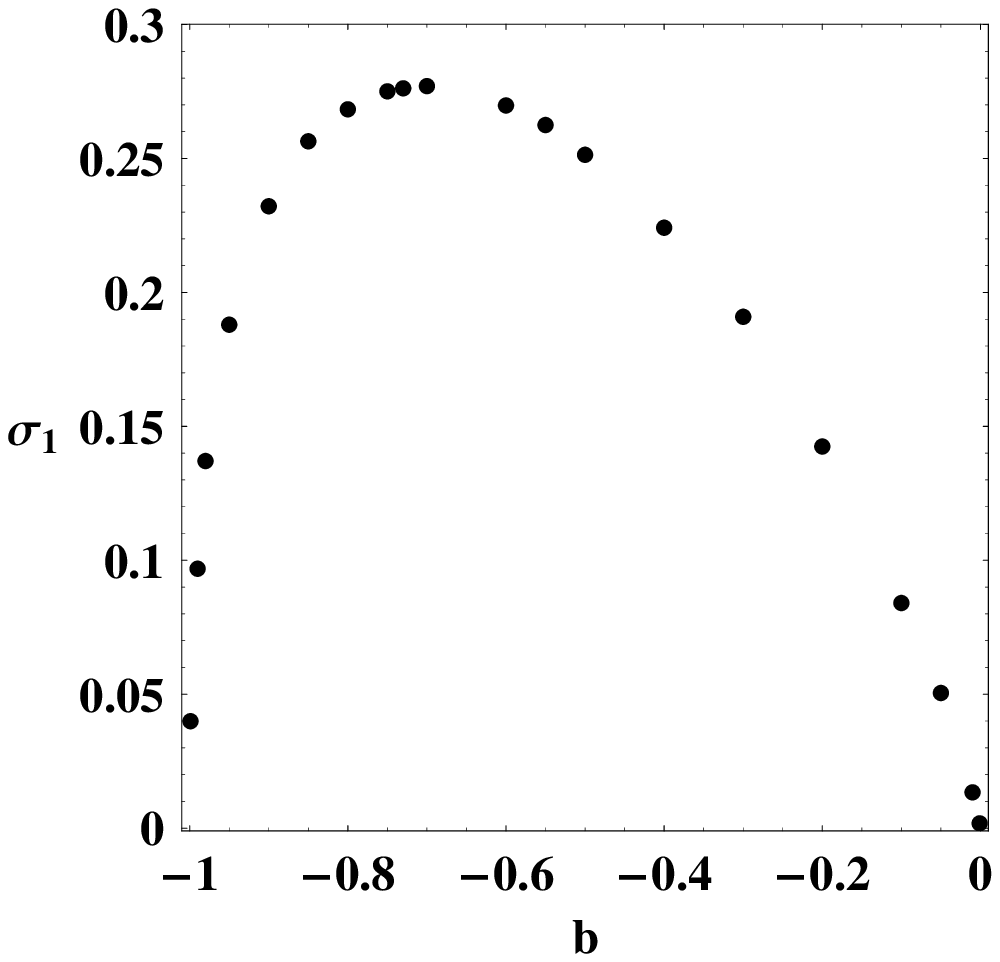}
\hfill
\includegraphics[scale=0.6,angle=0]{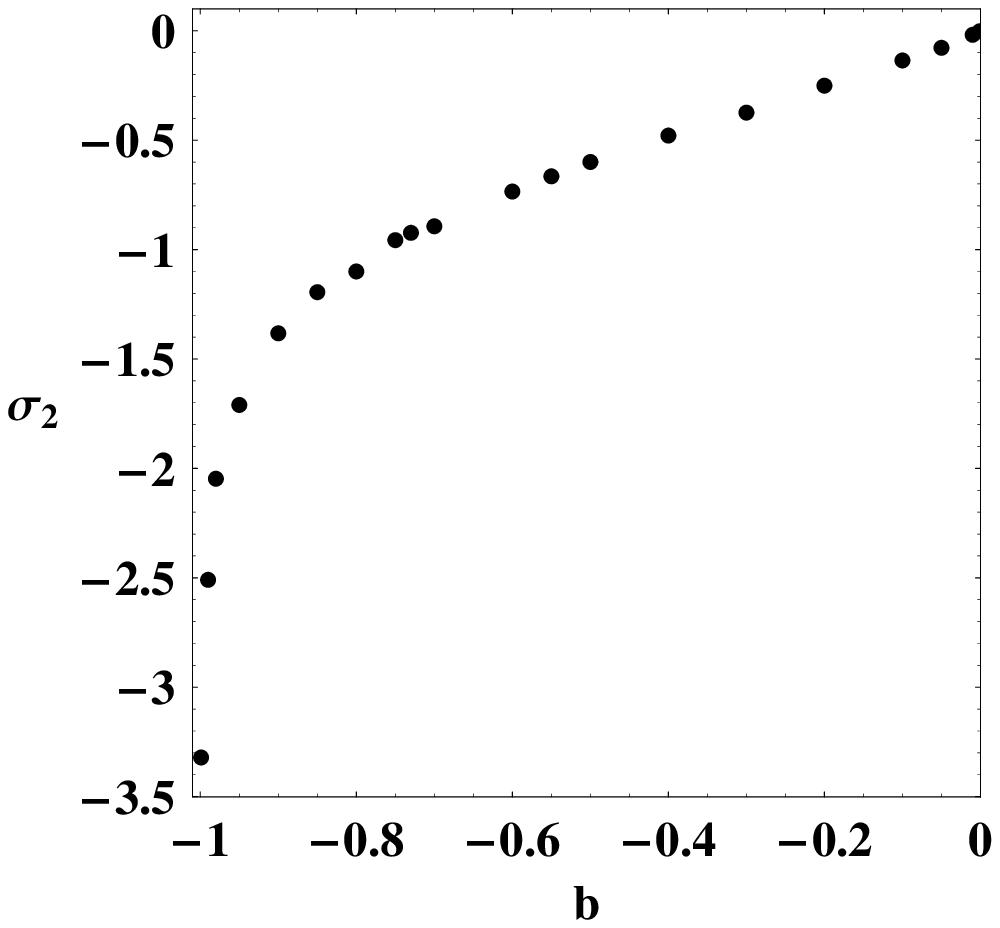}
\end{center}
\vskip .1cm
\protect\caption{Positive and negative
 Lyapunov exponents versus the parameter $b$}
\label{fig2a2b}
\end{figure}

The Lyapunov exponents for $ b \in ]-1, \, 0[$ are 
shown in Fig. \ref{fig2a2b}.
The largest Lyapunov exponent being
 {\em positive}, the attractor is chaotic.

Kaplan and Yorke~\cite{kap-yor-79} have conjectured that the dimension
of an attractor can be approximated from the spectrum of Lyapunov
exponents.
Essentially, this conjecture corresponds to plotting the sum
of Lyapunov exponents versus $n$ (the number of Lyapunov exponents, i.e.
the dimension of the system), the dimension being established by
finding where the curve intercepts the $n-$axis by linear
interpolation\footnote[1]{Note the comparison made in~\cite{chl-spr-04} 
for the correlation and Lyapunov dimensions using, for the latter, 
a polynomial interpolation instead of a linear one.}.
For a two-dimensional mapping, this gives
\begin{eqnarray}
\label{dimKY}
D_{KY} \,=\,\,\, \,  1\,\, -{\frac{\sigma_1}{\sigma_2}}
\end{eqnarray}
where $\sigma_1$ and $\sigma_2$ are respectively, the positive and
negative Lyapunov exponents. This dimension is
 expected to be close to
other dimensions such as box-counting, information and correlation
dimension~\cite{gra-pro-83} for typical strange attractors.
Kaplan-Yorke dimension (also called Lyapunov dimension) has been shown to
identify with the information
dimension for Baker's transformation. It has been tested for H\'enon
mapping~\cite{ru-ha-ot-80}.

\begin{figure}
\begin{center}
\includegraphics[scale=0.7,angle=0]{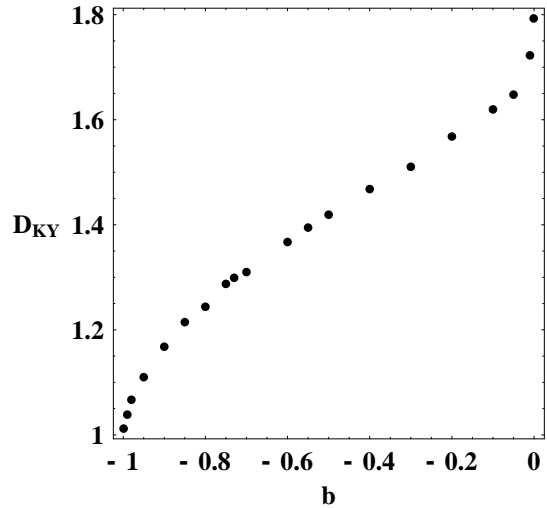}
\end{center}
\vskip .1cm
\protect\caption{Kaplan-Yorke (or Lyapunov) 
dimension versus the parameter $b$}
\label{fig3}
\end{figure}

Using Kaplan-Yorke conjecture, we can compute the (fractal) dimension of the
attractor which is shown in Fig.\ref{fig3}. For $b \in \,  ]-1, \, 0[$ the
dimension of the attractor corresponds to fractals. The attractor is thus
 a {\em strange attractor}.

The dimension obtained from the Lyapunov exponents is given from a
conjecture. To be more convincing 
on the fractal nature of the attractor,
we have calculated the (fractal) dimension by the
 box-counting method for some values of $b$.
Box-counting dimension is believed to coincide, for most systems,
with Hausdorf-Besicovitch dimension.
The box-counting dimension amounts to
 considering boxes of side $\epsilon$
covering the attractor, then counting the number of boxes $N(\epsilon)$
necessary to contain all the points.
The box-counting dimension is the limit as
 $ \,\epsilon \, \rightarrow \, 0 \,$ of
\begin{eqnarray}
D_{box}\,  =\, \,  \,  \, 
 - {\frac{\ln \left( N(\epsilon) \right)}{\ln (\epsilon)}}.
\end{eqnarray}

The calculations are carried out in the variables $\theta_u$,\,  $\theta_v$
(which are in one-to-one correspondence with $u,\,v$).
The results given in Table \ref{tab1}, 
show an agreement with the dimension
computed from the Lyapunov exponents as far as the fractal nature is
concerned. Note that for this mapping,
the Lyapunov (Kaplan-Yorke) dimension is less than  
the box-counting dimension (and is a lower bound~\cite{kap-yor-79}).

\begin{table}
\begin{center}
\begin{tabular}{|l|l|l|l|l|l|l|l|}
\hline
$b$& -0.9 & -0.8 & -0.6 & -0.5 & -0.4 & -0.3 & -0.2     \\ 
\hline
$D_{KY}$  & 1.17  & 1.24 & 1.37 & 1.42 & 1.47 & 1.51 & 1.57     \\ 
$D_{box}$ & 1.34  & 1.36 & 1.44 & 1.52 & 1.52 & 1.56 & 1.66     \\ 
\hline
\end{tabular}
\end{center}
\vskip 0.2cm
\protect\caption{Kaplan-York and 
Box-counting dimensions for some values
of the parameter $b$}
\label{tab1}
\end{table}

\vskip .1cm

{\bf Remark:}
The simplicity of the birational mapping $K$ makes it a good working example of
many tests.
For instance, it will be straightforward to
compute the Lyapunov exponents and the fractal dimension from the knowledge
of the first fixed points~\cite{cvitanovic-88}.
How many fixed points will be needed for that
purpose? Are the fixed points sufficient to completely  characterize
the mapping in some ergodic analysis?

\section{Conclusion}
\label{concl}

In this paper we have, first,  considered four birational mappings.
Three of them ($K_1$, $K_2$, $K_4$) have been already analyzed in
previous papers and the fourth mapping ($K_3$) is taken from the literature
 on strongly regular graphs~\cite{jaeger-92}.
For these mappings, we have considered their
{\em post critical set versus the occurrence of covariant curves and
preserved meromorphic two-form}.
In these working examples, the post-critical set is ``long'' in both
directions (forwards and backwards).
We have shown that the knowledge of the orbits of the critical set
allows to obtain the algebraic covariant curves.

We have built a {\em birational deformation} of the H\'enon map, $H_d$,
having a ``short'' post-critical set.
This birational mapping shows a continuous deformation of the original
H\'enon {\em strange attractor}.
For the values of the parameters (giving a strange attractor for $H_d$),
the backward mapping $H_d^{-1}$ shows an unbounded attracting set contrary
to the backward H\'enon map that gives divergent orbits.

We have focused on a birational mapping $K$ (slight modification
 of a birational mapping of Bedford and Kim~\cite{bed-kim-05})
 which has also a ``short''
post-critical set, with probably no preserved meromorphic
 two-forms (in view of the non standard fixed points).   
The mapping shows an attracting set, which 
 {\em passes the tests} of being a
strange attractor.

In view of these examples, we saw
that  birational mappings with ``short'' post-critical set
had no covariant curves. We also saw
that strange attractors may occur for birational mappings
with ``short'' post-critical set and they are not necessarily confined.

\ack
SH thanks the hospitality of LPTMC, where part of this
work has been completed. 
This work has been performed without any support
 of the ANR, the ERC, or the MAE.

\appendix

\section{The birational mapping $K_3$}
\label{K3}

We consider a $\, 3 \times 3\, $  matrix, acting on 
the three homogeneous variables $\, (x,\, y,\, z)\,$, taken from
the analyzis of strongly regular graphs~\cite{jaeger-92}
matrix
\begin{equation}
\label{theM}
M \, = \, \, 
\left [\begin {array}{ccc} 
2 &  a & b\\
2 & -1+c & -1-c\\
2 & -1-d & -1+d
\end {array}\right ],
\end{equation}
and the associated collineation  $\, C$. denoting $j$ the Cremona inverse
(\ref{thej}),
the mapping $K_3=\,C^{-1}\cdot j \cdot C \cdot j$ 
depends on four parameters. Here we fix $a=\,6,\, b=\,6,\, d=\,c$.
The birational mapping in the variables $u=\, x/z$, $v=\,y/z$
is given by (\ref{K3birat}). 
The Jacobian\footnote[5]{Note that the mapping depends on $c^2$.
However, we do not rename the parameter $c^2$ for easy
presentation, see the factorized expression
 (\ref{JK4}) in the Jacobian.}
reads:
\begin{eqnarray}
\label{JK4}
&& J[K_3](u, \, v)\, =\,\,\, \,  
- \, 5488 \, {\frac {v \, N_{uv} }{u^2\, D_{uv}^3 }} \cdot  
 \left((c+1)\, uv\, -(c-1)\, u\, -2v \right) \, 
\nonumber \\
&& \qquad \qquad \times \, 
  ((c-1)\cdot  uv\,\, -(c+1)\, u\,\, +2v).   
\end{eqnarray}

The exceptional varieties of the mapping are: 
\begin{eqnarray}
&& {\cal E}(K_3) \, = \, \, \, 
 \left \{ V_1, \, V_2, \, V_3, \, V_4, \, V_5, \, V_6 \right \}
 \nonumber \\
&&\qquad  = \left\{
\begin{array}{ll}
  (u=0); \,\, (v= \,{-3u \over 1+3u});  \,(v=0);\,\,
(v= \,{\frac{u(c+1)}{cu-u+2}}); \nonumber \\
(v \,=  \,\,{\frac{u(c-1)}{cu+u-2}}); \, \, \,  \,(D_{uv} \,= \, \,0) 
\end{array}    \right\}.
\end{eqnarray}

The post critical set for $V_1$ is variable dependent.
The orbit for $V_3$ reads
\begin{eqnarray}
&& K_3^n(V_3) \, = \, \, \,\,
\Bigl({1 \over 2} \left(1+(-1)^n \right)\,\,
+{1 \over 2} \left(1-(-1)^n \right)\, v_n, \,\, \, v_n  \Bigr), \quad
\nonumber \\
&& \qquad  v_n \, =\,\,\, \, {\frac{f(c)-f(-c)}{g(c)-g(-c)}}, 
\nonumber \\
&& f(c)\,=\,\, \,(c-7) \cdot \left( c-21\,\, - (-1)^n (3c-7) \right)
\cdot (c+1)^n, \nonumber \\
&& g(c)\,=\,\, \,(c-7) \cdot \left( c+21\,\, - (-1)^n (3c+7) \right) 
\cdot (c-1)^n. \nonumber
\end{eqnarray}

The orbit for $V_2$, after
 $(1, \,1) \rightarrow (\infty, \, \infty)$, gives
similar expression as $V_3$ and reads ($n \ge 3$):
\begin{eqnarray}
&& K_3^n(V_2) \, = \, \,\, \Bigl({1 \over 2} \left(1-(-1)^n \right)
+{1 \over 2} \left(1+(-1)^n \right)\, v_n, \,\, v_n  \Bigr), \quad 
\nonumber \\
&& \qquad  v_n =\,\, \,  {\frac{f(c)+f(-c)}{g(c)+g(-c)}}, \nonumber \\
&& f(c) \,=\, \, (c-7) \cdot \left( 3c+71 \,\, + (-1)^n (c+21) \right)
\cdot (c+1)^{n-2}, \nonumber \\
&& g(c) \,=\, \, \, (c-7)\cdot  \left( 3c-7 \, \, + (-1)^n (c-21) \right) 
\cdot (c-1)^{n-2}. \nonumber
\end{eqnarray}

The orbit for $V_5$ is identical, with the change 
$c \, \rightarrow \,  -c$, with
the orbit for $V_4$ which reads:
\begin{eqnarray}
&& K_3^n(V_4) \, = \,\,\,  \Bigl({\frac{(c-1)(c+7)^n\,-(c+1)(c-7)^n}
{ f(c) \cdot v_n\, + (-1)^n \, f(-c)}} \cdot v_n ,\,\, v_n \Bigr),
 \nonumber \\
&& \qquad  v_n \,  =\,\, \,
 {1\, \,  \over 2} \left(1-(-1)^n \right) \cdot {c+7 \over c-7}\,\,
 + {1 \over 2} \left(1+(-1)^n \right) \cdot {c-7 \over c+7},
 \nonumber \\
&& f(c)\, =\, \, 
(c-1)\cdot  (c + (-1)^n\,7) \cdot (c+7)^{n-1}. \nonumber 
\end{eqnarray}

Finally, the post critical set $K_3^n(V_6)$
 reads (after $(\infty, \infty)$,
this is identical to $K_3^n(V_3)$ with $c \, \rightarrow \, -c$
 and a shift in $n$):
\begin{eqnarray}
&& K_3^n(V_6)\,  = \, \,  \,
\Bigl({1 \over 2} \, (1+(-1)^n) \, \, \,      +{1 \over 2} 
\left(1-(-1)^n \right)\cdot  v_n, \,\, v_n  \Bigr), 
\quad \nonumber \\
&&\quad \quad v_n \, =\, \,\,\,
 {\frac{f(c)\, -f(-c)}{f(c)\, g(c)\,\, -f(-c)\, g(-c)}},
 \nonumber \\
&& f(c)\,=\,\,(c-7) \cdot \left( 3c+7 +(c+21) (-1)^n \right)
\cdot (c+1)^{n+1}, \nonumber \\
&& g(c)\,=\,\, {1 \over 2}\,  (1-(-1)^n) \cdot {c+7 \over c-7}
\,\,\,
 + {1 \over 2} \, (1+(-1)^n) \cdot {c-7 \over c+7}, \nonumber 
\end{eqnarray}

From these orbits, one finds easily the covariant curves.
The post critical set of $K_3^n(V_3)$ gives $u=\,1$ for $n$ even and
$v=\, u$ for $n$ odd, the covariant curve is thus $(u-1)(v-u)=\, 0$.
The post critical set of $K_3^n(V_4)$ gives
 $v\, -(c-7)/(c+7)=\, 0$ for $n$
even and $v\,-(c+7)/(c-7)=\,0$ for $n$ odd, leading to the covariant
curve $C_2=\, \left( (c+7)\cdot v\, -c+7 \right)\, 
\left( (c-7)\cdot v\, -c-7 \right)=\, 0$.
The post critical set of $K_3^n(V_6)$ gives the same covariant curve
as $K_3^n(V_3)$.

These curves are covariant but, alone, they do not construct a preserved
meromorphic two-form (see (\ref{preserv})). One obtains:
\begin{eqnarray}
&& m(u,v)\, =\, \, 
(u-1)\, (v-u)\cdot ((c+7)\, v\, -c+7)\cdot 
\left( (c-7)\, v\, -c-7 \right),
 \nonumber \\
&& {\frac{ m(u', v')}{m(u, v)}}\, 
 = \, \,\,\, 28\,\, {\frac{N_{uv}}{D_{uv}}}
\cdot J[K_3](u, \, v).    \nonumber
\end{eqnarray}

At this point, the mapping $K_3$ does not seem to 
 have a preserved meromorphic two-form.

However, if a meromorphic two-form exists, we know~\cite{bo-bo-ha-ma-05}
 that the fixed points of
 the mapping for which the Jacobian
is $J\, \ne\,  1$ should be located on a covariant curve corresponding to the 
meromorphic two-form. For this mapping, there are four
 fixed points of order one where $J\, =\, 1$
and two fixed points of order one where $J \, \ne \, 1$.
The latter read $(u=\, -5/2 \pm \sqrt{21}/2, \, v=\, 1)$ and 
are neither on $(u-1)(v-u)=\, 0$ nor on $C_2=\, 0$.
The line $v-1=\, 0$ should be covariant as this appears clearly from
the expression (\ref{theK4}) of the birational mapping $K_3$.

Producing this line $v-1=\, 0$ from the iterates 
of $V_1$ may call for a tricky
analysis. Instead, and since this is equivalent, we consider the backward
mapping and its ``long'' post critical set.
Cancelling the Jacobian (or its inverse) of $K_3^{-1}$, one obtains (among
others) the algebraic curve:
\begin{eqnarray}
28\,{u}^{2} \, +56\, \left(1+v \right)\, u \, 
- \left( {c}^{2}+35 \right) \, v^{2}\, 
+2\, ({c}^{2}-35)\,  v\,\,  -{c}^{2}\, -35\,=\,\, 0. \nonumber
\end{eqnarray}
Eliminating the variable $u$ between this curve and the iterates
$K_3^{-n}(u,v)$, will give the line $v-1=\,0$ common to both components.

Combining the covariant curves $(u-1)(v-u)=\, 0$, $C_2=\, 0$
 and the new line $v-1=\, 0$, one obtains
\begin{eqnarray}
&& m_3(u,v)\, =\, \,\, \, 
{\frac{ (u-1)(v-u)\cdot }{v-1}}
\cdot
 \left((c+7)\, v\, -c+7 \right) \, 
\left( (c-7)\,v\, -c-7 \right), \nonumber \\
&& {\frac{ m_3(u', v')}{m_3(u, v)}}\,\,
  =  \, \,\,\, J[K_3](u, \, v).
    \nonumber
\end{eqnarray}
giving the corresponding meromorphic two-form $du\, dv/m_3(u, \, v)$.

\section{Another birational mapping: $K_5$}
\label{K5}
We consider the collineation $C$ corresponding to matrix (\ref{theM})
but with the mapping constructed as $K_5\,=\,\,C \cdot j$.
This mapping arises in~\cite{be-ma-vi-91} and 
was considered in~\cite{bed-kim-04,viallet-96}.
For the values of the parameters $a=\,b=\, -1+q^2$, $c=\,d=\,q$,
 it was shown~\cite{be-ma-vi-91} that it has 
an algebraic invariant for all values
of the parameter $q$.
Here, we take the parameters as $a=\,b=\,q^2$,
 $c=\,d=\,q$ and the birational mapping
is non integrable for generic values of $q$.
The birational mapping reads
\begin{eqnarray}
&& K_5: \,\, \quad (u,v)\quad
\,\,\longrightarrow \,\, \qquad (u', v')\, =\,\,  \nonumber\\
&& \quad \quad \qquad \, =\,\,\,
\Bigl( {\frac{q^2 \cdot  (1+v)\,u  + 2v}{D_{uv}}},\,\,\,
1- {\frac{2 q \cdot (v-1)\,u }{D_{uv} }} \Bigr),  \\
&& D_{uv}\,=\,\, \,
(q-1)\cdot  uv \,\, -(q+1)\cdot u\, + 2v,  \nonumber
\end{eqnarray}
with Jacobian:
\begin{eqnarray}
&& J[K_5](u, \, v)\, =\,\, \,\,  - 8 \,  (1\,+q^2)\cdot  q
\cdot {\frac { u v }{ D_{uv}^3 }}. 
\end{eqnarray}

Its critical set reads
\begin{eqnarray}
{\cal E}(K_5) \, = \,\,\,\,   \left \{ V_1,\, V_2, \,V_3 \right \}
\,\,=\,\,\,\, 
\left \{ (v=0);\,\,  (D_{uv}=0);\, \, (u=0) \right \},  
\end{eqnarray}
and the first iterates of the critical set are given by
\begin{eqnarray}
&&K_5^n(V_1) \,=\,\, \Bigl( -{\frac{q^2}{q+1}}, \, {\frac{1-q}{q+1}}\Bigr)
\qquad \longrightarrow \qquad 
\Bigl( {\frac{1-q^2-q^4}{1+q^4}}, \, {\frac{1-q^4}{1+q^4}}\Bigr),
 \nonumber \\
&&K_5^n(V_2)\, = \,\,(\infty, \infty) \,\, \, \rightarrow  \,\, \,
\Bigl( {\frac{q^2}{q-1}}, \, {\frac{1+q}{1-q}}\Bigr)
\,\, \,  \rightarrow  \,\, \,
\Bigl( {\frac{1-q^2-q^4}{1-q^4}}, \, {\frac{1+q^4}{1-q^4}}\Bigr).
 \nonumber 
\end{eqnarray}
Those of $K_5^n(V_3)$ are the same as $K_5^n(V_2)$ after blowing down first
on the point $(1, 1)$.

The post-critical set for the {\em backward} mapping is also ``long'' and the
orbits have similar expressions. 
The post-critical set, for both forward and
backward mapping, is ``long''.
The degree growth in the parameter $q$ of the  {\em iterates of the
critical set} being exponential, there is no preserved meromorphic two-form.
The phase portraits of this mapping show a foliation in the plane with
infinity of leaves, similar to mapping analyzed in~\cite{bo-bo-ha-ma-05}.
Note that, the line $v-1=\, 0$ is covariant as
easily seen from the expression of $K_5$.
The phase portrait, shows however, no accumulation of points near this line.

\section{A parameter-free birational mapping}
\label{K6}
This mapping is taken from~\cite{bo-ha-ma-97} and originates from
lattice statistical mechanics and is related to mapping $K_1$.
It is  parameter-free, non integrable and reads
\begin{eqnarray}
  K_6 : \quad (u, v)\,
\quad  \,\longrightarrow \,\,\quad  (u', v')
 \, \, = \,\, \, \,   \Bigl(v, \,\, {\frac {1+v-uv}{u v}} \Bigr).  
\end{eqnarray}
Its Jacobian reads:
\begin{eqnarray}
  J[K_6] \, \, =\,\,\,\, \, {\frac {1+v}{u^2 v}}.  
\end{eqnarray}
The orbits of the critical set read:
\begin{eqnarray}
&& K_6^n(v=-1)\, = \,\, \,  (-1, \, -1)
 \quad  \longrightarrow  \quad \left(-1, \, -1 \right),
 \nonumber \\
&& K_6^n(u=0)\, =\,\,
  (v, \, \infty) \rightarrow  (\infty, \, (1-v)/v)
\rightarrow  ((1-v)/v, \, -1) \rightarrow  (-1, \, -1), 
\nonumber \\
&& K_6^n(v=0)\, = \,\, (0, \infty) 
\quad \rightarrow \quad  (\infty,  \,\infty) 
\quad  \rightarrow \quad  (\infty, \, -1) 
\quad \rightarrow \quad  (-1, \, -1).
 \nonumber 
\end{eqnarray}
The post-critical set is ``short'' and there is an 
attracting set which is the point
$(-1, \, -1)$.

For the backward mapping, the orbits of the critical are
\begin{eqnarray}
&& K_6^{-n}(u=-1) =\,  (0, \, -1)
\quad \rightarrow \quad  (\infty, \, 0)
 \quad \rightarrow \quad  (1, \, \infty)
 \quad \rightarrow \quad   (0, \, 1),
 \nonumber \\
&& K_6^{-n}(u=0) =\, (\infty, \, 0) \quad 
 \rightarrow \quad  (1, \infty)
 \quad \rightarrow  \quad  (0, \, 1)  \quad \rightarrow
 \quad  (\infty, \, 0), \nonumber \\
&& K_6^{-n}(v=-1) = \, (\infty, \, u) \quad 
 \rightarrow \quad  (1/(1+u),  \,\infty) \quad 
\rightarrow \quad  (0, \, 1/(1+u)) \nonumber \\
&& \quad \quad \longrightarrow  \quad (\infty, \,  \,0)
\quad \longrightarrow \quad  (1, \, \infty) \quad 
 \longrightarrow \quad  (0, \, 1). \nonumber
\end{eqnarray}
The post critical set is short and there is an attracting set which is the
cycle of order three,
$\left(\infty, \, 0 \right) \rightarrow (1, \, \infty)
 \rightarrow \left(0, \, 1 \right)$.

Note that one may remark (see the form of the mapping) that $v+1=0$ is
covariant for the forward mapping, but it is not covariant for the
backward mapping, where it gives birth
 to an attracting point of order three.

The Jacobian evaluated at the successive fixed points gives the
following. There is one fixed point of order one
 with\footnote[1]{For the backward mapping the value of the
 Jacobian evaluated at the fixed point is the inverse
 of the value corresponding to the forward mapping.} $J=\,2$.
There is no fixed point of order three. There is only one fixed
point for the orders $2,\, 4,\, 5,\, 6$ with respectively
$J=\,\, 2,\, 2,\, -5,\, 1$.

\end{document}